\documentclass[twocolumn,prl,aps]{revtex4}
\usepackage{amsbsy}
\usepackage{latexsym,epsfig,graphicx}
\usepackage{dcolumn}
\usepackage{graphicx}
\usepackage{epstopdf}
\usepackage[T1]{fontenc}
\usepackage{textcomp}
\usepackage{subfigure}
\usepackage{comment}
\usepackage{bm}
\usepackage{mathrsfs}
\usepackage{braket}
\usepackage{amsfonts}
\usepackage{amsmath}
\usepackage{color}
\usepackage{amssymb}
\usepackage{xspace}
\usepackage{tabularx}
\usepackage{longtable}
\usepackage[normalem]{ulem}
\usepackage[colorlinks=true, letterpaper=true, pdfstartview=FitV, linkcolor=blue, citecolor=blue, urlcolor=blue]{hyperref}
\begin{document}
\title{Optimal control of population transfer in multi-level systems by dynamical quantum geometric tensor
\footnote{\par\vspace{0pt}
\noindent The paper is an English translated version of the original Chinese paper published in Acta Physica Sinica. Please cite the paper as: LI Guanqiang, ZHANG Yuqi, GUO Hao, DONG Youjiao, LIN Zhiyu, PENG Ping. Optimal control of population transfer in multi-level systems by dynamical quantum geometric tensor. Acta Phys. Sin., 2025, 74(10): 100304.
\\
\noindent DOI: \href{https://wulixb.iphy.ac.cn/article/doi/10.7498/aps.74.20250210}{10.7498/aps.74.20250210}
\\
\noindent CSTR: \href{https://wulixb.iphy.ac.cn/article/doi/10.7498/aps.74.20250210}{32037.14.aps.74.20250210}
\par\vspace{0pt}}}
\author{LI Guanqiang }\email[Corresponding author:~]{liguanqiang@sust.edu.cn. }
\author{ZHANG Yuqi, GUO Hao, DONG Youjiao, LIN Zhiyu, and PENG Ping}
\affiliation{School of Physics and Information Science, Shaanxi University of Science and Technology, Xi'an 710021, China}

\begin{abstract}
The optimal control of population transfer for multi-level systems is investigated from the perspective of quantum geometry. Firstly, the general theoretical framework of optimizing the stimulated Raman adiabatic passage (STIRAP) scheme based on the dynamical quantum geometric tensor is given, and then the dynamical quantum geometric tensor and the nonadiabatic transition rate are calculated by taking the detuned $\Lambda$-type three-level system and tripod-type four-level system for example. Secondly, the transfer dynamics of the particle population of the system are investigated in detail. For a three-level system, the optimal STIRAP scheme has an efficiency of over 98\% in transferring the population to the final state, while the transfer efficiency of traditional STIRAP is about 72\%. The superposition states with arbitrary proportions can be efficiently prepared for a four-level system due to the decoupling of the degenerate dark states. Finally, the influences of system parameters, such as the operation time of the Rabi pulses, the amplitude fluctuation and the single-photon detuning, on the transfer process are discussed. Especially, the phenomenon of the adiabatic resonance transfer is revealed. Choosing the pulse parameters in the resonance window can reduce the infidelity of the population transfer to below $10^{-3}$. It is found that the optimal STIRAP scheme by the dynamical quantum geometric tensor provides faster and more efficient transfer than the traditional STIRAP scheme.

\end{abstract}

\maketitle

\section*{I. Introduction}
Quantum adiabatic control technology provides a set of effective ideas to control quantum systems, and its theoretical basis is the quantum adiabatic theorem. In 1928, Born and Fock proposed the theory that a quantum system evolves adiabatically following its energy eigenstate~\cite{MBorn1928, ZWu2005}. They proved that if there is an energy gap between a particular energy eigenvalue of the system and the rest of the energy spectrum, and a given time-dependent perturbation acting on it is slow enough, then the system can always remain in its corresponding instantaneous eigenstate. The quantum adiabatic theorem is one of the most important conclusions in quantum theory, which has a wide range of applications in theory and practice~\cite{SBarry1983, DXiao2010}. In quantum adiabatic evolution, the process of stimulated Raman adiabatic passage (STIRAP) is one of the most effective schemes to realize population transfer~\cite{NVV2017}. In 1991, Shore et al. demonstrated an efficient method for adiabatic population transfer between two discrete quantum states in atoms or molecules~\cite{BWS1991}. The method was soon extended to systems with three discrete quantum states~\cite{BWS2013, XPSun2007, SYMeng2009, GQLi2011, GQLi2012}. Two delayed laser pulses are applied to a three-level system to achieve complete population transfer between two lower energy levels (via a higher intermediate level). In particular, the pulse sequence employed in the STIRAP scheme is counter-intuitive, i.e., the Stokes laser pulse couples the intermediate and final states, switching on prior to (but overlapping with) the pump laser pulse, which couples the initial and intermediate states. The intensity of the laser field should be high enough to produce multiple Rabi oscillation cycles. The laser-induced coherence between quantum states can be adjusted by time delay to ensure that the instantaneous population of the intermediate state is almost zero and the population loss caused by radiation attenuation is avoided~\cite{NVV2017, BWS1991, KBer2015}. In order to achieve a more efficient transfer, this requires the system to meet specific adiabatic conditions~\cite{MPFewell1997}. However, the process that satisfies the adiabatic condition must be an extremely slow process, which usually has significant disadvantages in practical applications.

In order to speed up the quantum adiabatic process and maintain high transfer efficiency, researchers have proposed some new control schemes, such as quantum shortcut to adiabaticity (STA)~\cite{DGu2019,THa2024,XChen2010}. This scheme has been discussed in depth in many systems, such as nuclear magnetic resonance spin system~\cite{JMi2017}, quantum dot system~\cite{YBan2012}, atomic system in cavity~\cite{TOpa2016,LTian2012,SBar2013} and superconducting qubit system~\cite{XMYu2025}. In the manipulation of macroscopic quantum systems, nonadiabatic quantum control of spinor Bose-Einstein condensates has also been studied~\cite{SMa2012,SMa2016}. In addition, the current methods that can accelerate the quantum adiabatic conversion include the quantum driving without transition~\cite{MDem2003,MVBer2009}, the Lewis-Riesenfeld adiabatic invariant theory~\cite{HRLew1969,XChen2011}, the fast-forward method~\cite{SMas2015,SMas2008}, and the non-Hermitian shortcut technique~\cite{BTTor2014,BTTor2013,GQLi2017}. These methods lay a solid theoretical foundation for the adiabatic shortcut of population transfer process in quantum system. Recently, researchers have proposed an optimization method for the STIRAP based on the dynamical quantum geometric tensor, which uses optimal control techniques to implement the evolution of quantum states along the geodesic path in parameter space to eliminate or reduce the non-adiabatic effects in the process~\cite{KZLi2024,JFChen2022}. This method can significantly shorten the time required for the adiabatic quantum transfer process and is robust to systematic and random errors. However, the existing research is mainly focused on the non-degenerate three-level system with zero detuning, and the multi-level system with detuning and level degeneracy has not been studied.

The purpose of this paper is to study the optimization of the STIRAP in $\Lambda$-type three-level system and tripod-type four-level system with single-photon detuning by using the quantum geometric tensor. Sec.~II presents a general theoretical framework for optimal control of multi-level STIRAP technology based on the dynamical quantum geometric tensor. The optimization scheme and calculation results for a three-level system with single-photon detuning are presented in detail in Sec.~III. Sec.~IV presents the optimization scheme and calculation results for a four-level system with detuning, focusing on the decoupling of degenerate double dark states and the fast and efficient preparation of quantum superposition states in a four-level system. By comparing with the traditional STIRAP method, this paper demonstrates the robustness of the optimized STIRAP method in the process of quantum state population transfer of multi-level systems. Finally, the conclusion is given in Sec.~V.
	
\section*{II. General theory of optimizing STIRAP based on dynamical quantum geometric tensor}
First, we introduce the definition of the dynamical quantum geometric tensor and the general theoretical framework for optimizing STIRAP based on the tensor (hereinafter referred to as optimized STIRAP or OSTIRAP). An $N$-dimensional quantum system described by a time-dependent Hamiltonian $H(t)$ with nondegenerate instantaneous eigenvalues $E_n(t)$ and eigenstates $\ket{E_n(t)}$ (where $1 \leq n \leq N$) is considered. According to the quantum adiabatic theorem, if the system is initially in the $n$th eigenstate of the Hamiltonian, the final state of the system will remain in its $n$th adiabatic eigenstate as long as the parameters change slowly enough. However, for a process occurring within a finite time, the time evolution of the system will induce a nonadiabatic transition, leading to a deviation from the adiabatic eigenstate and ultimately affecting the population transfer efficiency of the system.

In order to determine the degree of nonadiabatic transition, the time-evolving state can be expanded in the adiabatic basis, that is,
$\ket{\psi_n(t)} = \sum_{l} c_{nl}(t) \ket{E_l(t)}$.
The Schr\"{o}dinger equation satisfied by the state vector can be transformed into an equation for the amplitude of the state vector:
\begin{equation}
\frac { \partial c _ { n l } } { \partial t } = - i E _ { l } ( t )c _ { n l } - \sum _ { m }  \langle E _ { l } ( t ) | \frac { \partial } { \partial t } | E _ { m } ( t ) \rangle c _ { n m }. \label{Schrodinger equation}
\end{equation}
The degree of nonadiabatic transition of the adiabatic eigenstate $\ket{E_n(t)}$ can be quantitatively described by the nonadiabatic transition probability $P_n^T(t) = \sum_{l \neq n} |c_{nl}(t)|^2$. According to the higher-order adiabatic approximation~\cite{CPSun1988,GRig2008,JFChen2019}, there are upper and lower bounds for the first-order term of $P_n^T(t)$, that is,
$P_{n,-}^T(t) \leq P_n^T(t) \leq P_{n,+}^T(t)$, and
\begin{equation}
P _ { n , \pm } ^ { T } ( t ) = \frac { 1 } { \tau ^ { 2 } } \sum _ { l \neq n } \left[ | \tilde { T } _ { n l } ( \frac { t } { \tau } )| \pm | \tilde { T } _ { n l } ( 0 ) | \right] ^ { 2 }, \label{shangjie}
\end{equation}
where $\tau$ is the time required for the evolution process. $\tilde { T } _ { n l } $ describes the nonadiabatic transition rate between the $n$th and $l$th energy levels in the form of
\begin{equation}
\widetilde{T}_{nl}(s)=
\frac{
\langle E_l(s)\vert \tfrac{\partial}{\partial s} \vert E_n(s)\rangle
}{
E_n(s) - E_l(s)},
\quad
1 \le n,l \le N,
\end{equation}
where $s = t / \tau$ is the normalized time. The total nonadiabatic transition rate of the system is
$\widetilde{T}_n(s) \equiv [ \sum_{l \neq n} | \widetilde{T}_{nl}(s) |^2]^{1/2}$.
The Hamiltonian $H(t)$ of the system is time-dependent if the parameters $R_{p,q}(t)$ ($1 \leq p,q \leq M$) are time-dependent.
The total nonadiabatic transition rate can be calculated using the dynamical quantum geometric tensor~\cite{KZLi2024,JFChen2022}:
\begin{equation}
\widetilde{T}_n(s)=
	\left[
	\sum_{i,j}
	\mathrm{Re}\!\left(D_{n,pq}\right)
	\frac{\partial R_p}{\partial s}
	\frac{\partial R_q}{\partial s}
	\right]^{1/2}, \label{Formula4}
\end{equation}
where the dynamical quantum geometric tensor $D_{n,pq}$ is defined as
\begin{equation}
D _ { n , p q } = \sum _ { l \neq n } \frac { \langle { E } _ { l } | \frac { \partial H } { \partial { R } _ { p } } |  { E } _ { n } \rangle \langle { E } _ { n } | \frac { \partial H } { \partial R _ { q } } | { E } _ { l } \rangle } { \langle { E } _ { n } -{ E } _ { l } ) ^ { 4 } }. \label{DQGT}
\end{equation}
As a metric in the parameter space, $D_{n,pq}$ quantitatively characterizes the total nonadiabatic transition rate and plays a dominant role in the whole optimization process of the system~\cite{KZLi2024}. In order to keep the evolution process in the adiabatic passage, the global nonadiabatic transition rate $\widetilde{T}_n(s)$ should be as small as possible. This will make the time-evolving state deviate from the adiabatic eigenstate uniformly and leads to the adiabatic resonance phenomenon of the system, that is, the system returns to the adiabatic eigenstate at the resonance point~\cite{SOh2013}. At this time, the upper and lower bounds of the nonadiabatic transition probability in Eq.~(\ref{shangjie}) become $P_{n,\pm}^T = \pm 4 \widetilde{T}_n^2(s) / \tau^2$, which is inversely proportional to the square of the operation time. Based on an adiabatic control pulse with constant $\widetilde{T}_n(s)$, high efficiency quantum adiabatic population transfer can be achieved in a short time.

\section*{III. Optimized STIRAP scheme for a three-level system }
\subsection{A. Quantum geometric tensor of three-level system and calculation of nonadiabatic transition rate}
\begin{figure}[h]
	\centering
	\includegraphics[width=0.4\textwidth]{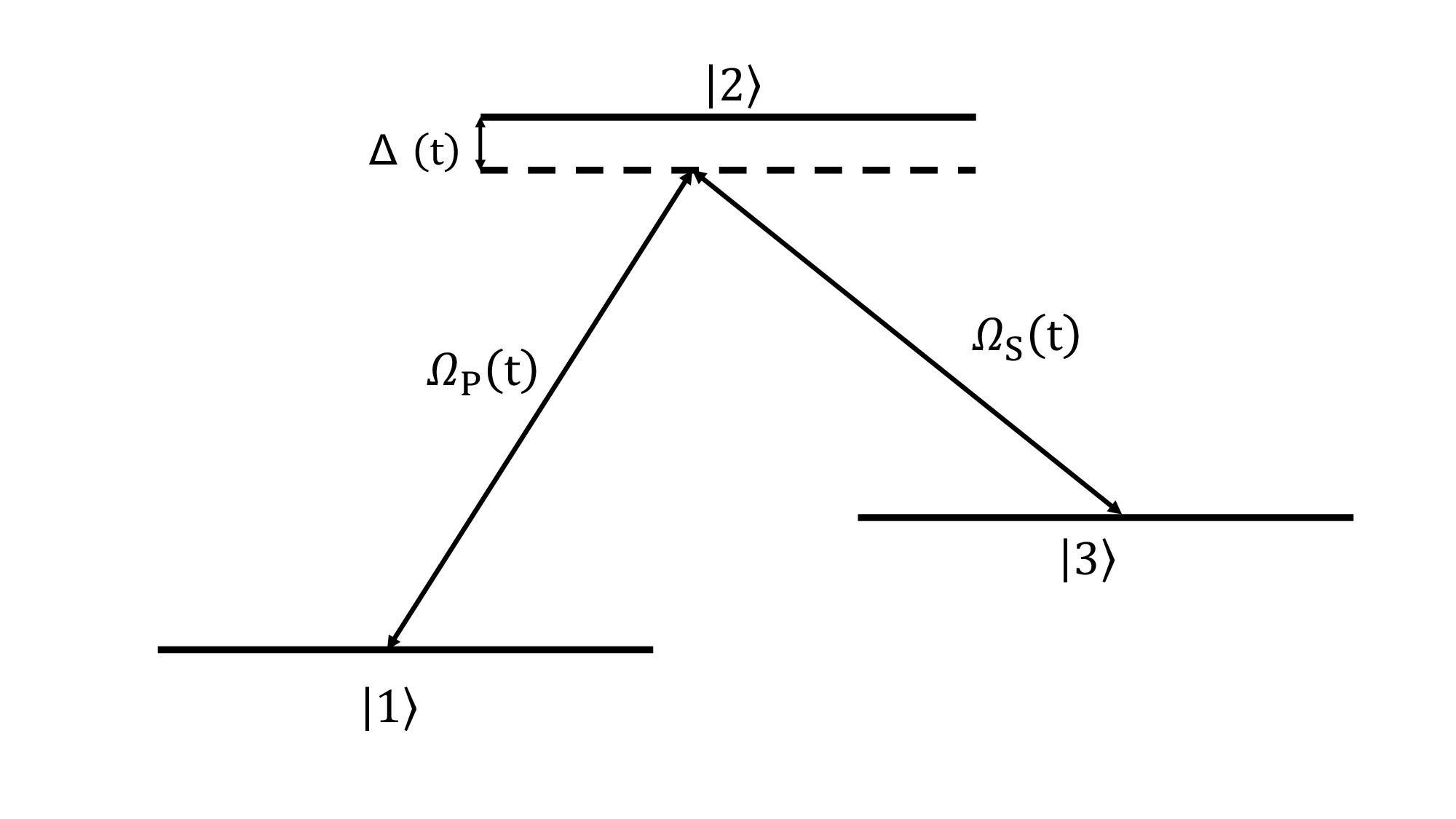}
	\caption{Schematic diagram of the STIRAP scheme for a three-level system with single photon detuning}
	\label{Figure1}
\end{figure}
The traditional STIRAP scheme of the $\Lambda$-type three-level system shown in Fig.~1 is optimized by using the above theory to realize the fast and efficient population transfer of the system from the state $\ket{1}$ to the state $\ket{3}$. In this paper, we focus on the optimization scheme and calculation results of a three-level system based on the dynamical quantum geometric tensor in the presence of single-photon detuning, which naturally includes the case of zero detuning.

In the rotating wave approximation, the Hamiltonian of the three-level system shown in Fig.~1 is~\cite{UGau1990,GRUnan2001,NVVit1998,NVVit2001,RUnan1998}:
\begin{equation}
	H(t)
	=
	\frac{\hbar}{2}
	\begin{pmatrix}
		0 & \Omega_P(t) & 0 \\
		\Omega_P^*(t) & \Delta(t) & \Omega_S(t) \\
		0 & \Omega_S^*(t) & 0
	\end{pmatrix},
\end{equation}
where $\Omega_{\mathrm{P}}(t)$ stands for the pump laser pulse coupling the states $\left|1\right\rangle$ and $\left|2\right\rangle$,
$\Omega_{\mathrm{S}}(t)$ stands for the Stokes laser pulse relating the states $\left|3\right\rangle$ and $\left|2\right\rangle$,
and $\Delta(t)$ is the single-photon detuning. The eigenvalues of this Hamiltonian are $\lambda_0 = 0$ and
$\lambda_\pm = \hbar\bigl[\Delta \pm \sqrt{\Delta^2 + \Omega^2}\bigr]/2$,
where $\Omega = \sqrt{\Omega_{\mathrm{S}}^2 + \Omega_{\mathrm{P}}^2}$.
The corresponding eigenstates of the system are
\begin{widetext}
	\begin{equation}
		\begin{aligned}
			\ket{\lambda_{+}} &=
			\sin\theta(t)\sin\phi(t)\ket{1}
			+ \cos\phi(t)\ket{2}
			+ \cos\theta(t)\sin\phi(t)\ket{3}, \\[3pt]
			\ket{\lambda_{0}} &=
			\cos\theta(t)\ket{1}
			- \sin\theta(t)\ket{3}, \\[3pt]
			\ket{\lambda_{-}} &=
			\sin\theta(t)\cos\phi(t)\ket{1}
			- \sin\phi(t)\ket{2}
			+ \cos\theta(t)\cos\phi(t)\ket{3},
		\end{aligned}
	\end{equation}
\end{widetext}
where the mixing angles satisfy
$\tan\theta(t) = \Omega_{\mathrm{P}}(t)/\Omega_{\mathrm{S}}(t)$ and $\tan 2\phi(t) = \Omega(t)/\Delta(t)$.
The eigenstate $\ket{\lambda_0}$ corresponding to the eigenvalue $\lambda_0 = 0$ is called the dark state,
and the remainders are called the bright states. According to Eq.~(\ref{DQGT}), the dynamical quantum geometric tensor of the three-level system with detuning can be calculated as follows~\cite{KZLi2024,JFChen2022}:
\begin{equation}
	\begin{aligned}
		D_{0,PP} &=
		\frac{4}{\hbar^{2}}
		\frac{\Omega_P^{2}}{\Omega^{2}}
		\frac{M_{+} + M_{-}}{M_{+}M_{-}}, \\[4pt]
		D_{0,SS} &=
		\frac{4}{\hbar^{2}}
		\frac{\Omega_S^{2}}{\Omega^{2}}
		\frac{M_{+} + M_{-}}{M_{+}M_{-}}, \\[4pt]
		D_{0,SP} &=
		-\frac{4}{\hbar^{2}}
		\frac{\Omega_S \Omega_P}{\Omega^{2}}
		\frac{M_{+} + M_{-}}{M_{+}M_{-}}
		= D_{0,PS},
	\end{aligned}
	\label{eq:D0_components}
\end{equation}
with
\begin{equation*}
	M_{\pm} =
	\bigl(\Delta \pm \sqrt{\Delta^{2} + \Omega^{2}}\bigr)^{2}
	\bigl[\Omega^{2} + \bigl(\Delta \pm \sqrt{\Delta^{2} + \Omega^{2}}\bigr)^{2}\bigr].
\end{equation*}
The total nonadiabatic transition rate of the three-level system can be obtained from Eq.~(\ref{Formula4}):
\begin{equation}
\tilde{T}_{n}(s) =
	\frac{2}{\hbar \Omega}
	\left[\bigl(\Omega_{S}^{2}\dot{\Omega}_{P}^{2}+\dot{\Omega}_{S}^{2}\Omega_{P}^{2}
	- 2\Omega_{S}\Omega_{P}\dot{\Omega}_{S}\dot{\Omega}_{P}\bigr)\frac{M_{+} + M_{-}}{M_{+} M_{-}}
	\right]^{1/2}. \label{eq:Tn}
\end{equation}
Eq.~(\ref{eq:Tn}) gives the total nonadiabatic transition rate of the system $\tilde{T}_n(s)$ in general cases. When $\Delta = 0$, the dynamical quantum geometric tensor and the total nonadiabatic transition rate of the system will reduce to the corresponding results without detuning~\cite{KZLi2024}. By keeping the total nonadiabatic transition rate constant, the nonadiabatic effects during the adiabatic evolution can be periodically self-cancelled, resulting in adiabatic transfer in a relatively short time~\cite{KZLi2024,SOh2013}.

The total nonadiabatic transition rate is used to optimize the pulses, which makes $\tilde{T}_n(s)$ a constant. A optimal combination of Rabi pulses is obtained:
\begin{equation}
	\begin{aligned}
{\widetilde{\Omega}}_P\left(s\right)=\Omega_0\sin{\left(\alpha s\right)},\ \ {\widetilde{\Omega}}_S\left(s\right)=\Omega_0\cos{\left(\alpha s\right)}.\label{RabiPulses}
	\end{aligned}
\end{equation}
The total nonadiabatic transition rate becomes
	\[
	\tilde{T}_{n}(s)
	= \frac{2\Omega_{0}}{\hbar^{2}}
	\Bigl(
	\Bigl\{[\Omega_{0}^{2} + (\Delta + \sqrt{\Delta^{2} + \Omega_{0}^{2}})^{2}]
	[(\Delta + \sqrt{\Delta^{2} + \Omega_{0}^{2}})^{2}]\Bigr\}^{-1}\]
	\[+\Bigl\{[\Omega_{0}^{2} + (\Delta - \sqrt{\Delta^{2} + \Omega_{0}^{2}})^{2}]
	[(\Delta - \sqrt{\Delta^{2} + \Omega_{0}^{2}})^{2}]\Bigr\}^{-1}
	\Bigr)^{1/2}.
	\]
According to this equation, increasing the intensity $\Omega_0$ of the laser pulse results in a reduced nonadiabatic transition rate and improved population transfer efficiency.
When $s = 0$, the evolved state of the system coincides with the dark state, and then it will evolve along the dark state channel. When $s = 1$, the system naturally evolves
to the final target state. It can be seen from Eq.~(\ref{RabiPulses}) that the phase $\alpha s$ of the optimized pulse will change from 0 to $\alpha$ in the range of $s = [0,1]$,
and $\alpha s$ is equivalent to the mixing angle $\theta = \arctan(\Omega_{\mathrm{P}} / \Omega_{\mathrm{S}})$ in the traditional STIRAP process.
In the following calculation, the optimized pulse parameter $\alpha = \pi / 2$ is taken, which enables the particles initially populated in the state $\ket{1}$ to be completely transferred to the state $\ket{3}$.

\begin{figure}[h]
\centering
\includegraphics[width=0.2\textwidth]{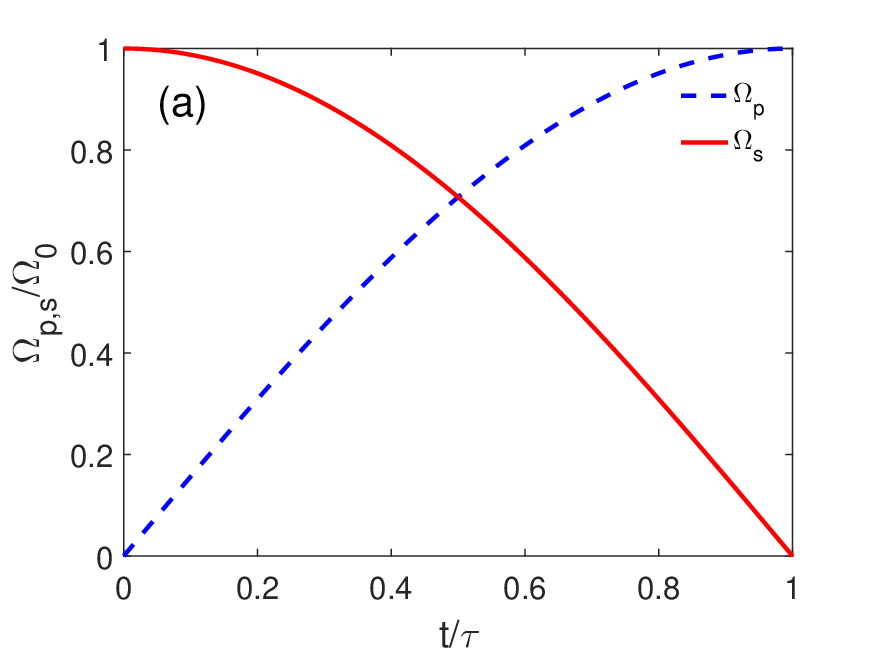}
\includegraphics[width=0.2\textwidth]{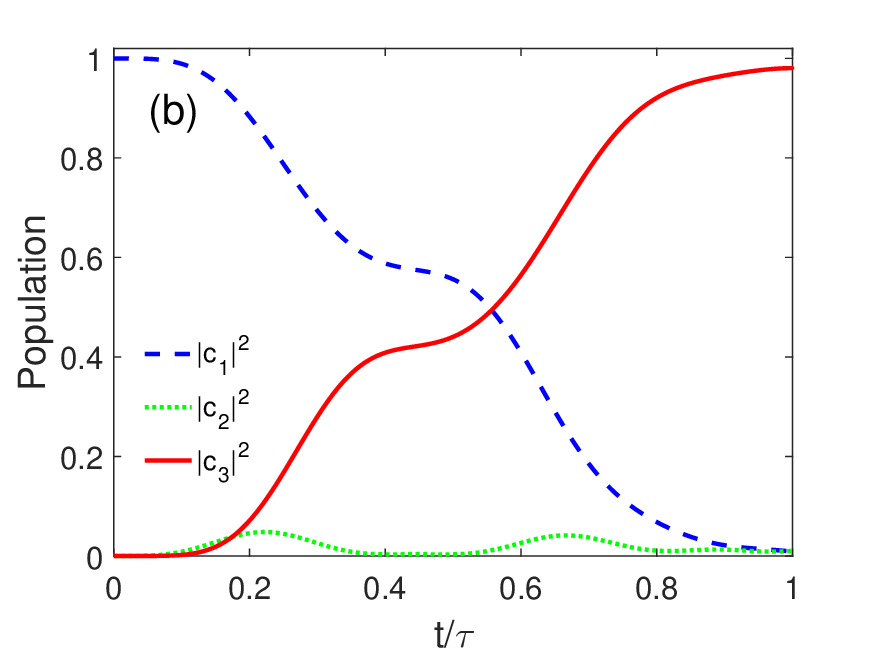}
\includegraphics[width=0.2\textwidth]{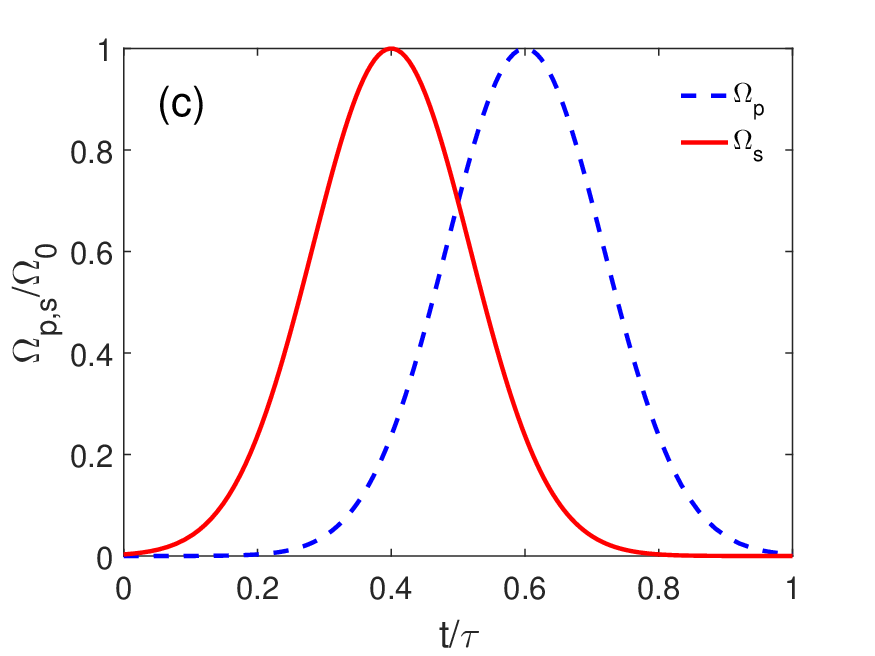}
\includegraphics[width=0.2\textwidth]{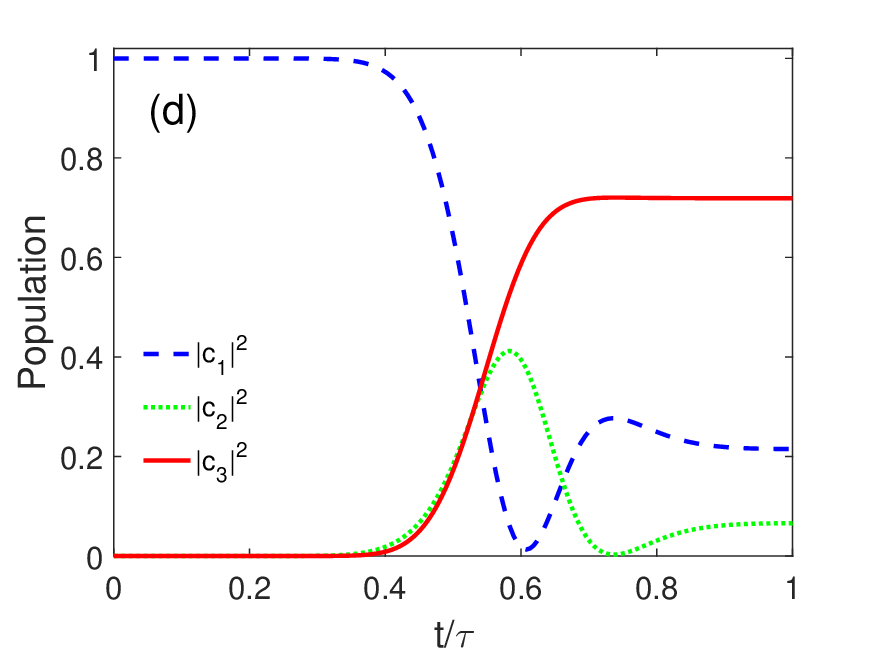}
\caption{Rabi pulse's structures and the evolution results of the populations for the three-level system: (a) Pulse structures for the optimal STIRAP; (b) Evolution of the populations for the optimal STIRAP; (c) Pulse structures for the traditional STIRAP; (d) Evolution of the populations for the traditional STIRAP (The pulse operating time $\tau=4\mu s$, the pulse peak $\Omega_0$=30.79MHz, and the detuning $\Delta=2\pi$MHz)}
\label{Figure2}
\end{figure}
\subsection{B. Population transfer dynamics of a three-level system}
Based on the Rabi pulses given in Eq.~(\ref{RabiPulses}), the population evolution of the system under the action of the optimized pulse can be calculated. The traditional STIRAP scheme uses the following Gaussian pulses:
\begin{equation}
	\begin{aligned}
\Omega_P\left(t\right)=\Omega_0\exp[-(t-\tau/2-\mu)^2/\sigma^2],
\\
\Omega_S\left(t\right)=\Omega_0\exp[-(t-\tau/2+\mu)^2/\sigma^2].
	\end{aligned}
\end{equation}
By comparing the evolution results of the optimized STIRAP and traditional STIRAP, the superiority of the optimized STIRAP using the dynamic quantum geometric tensor is demonstrated. In the calculation, the full width at half maximum of the  pulse $\sigma = \tau / 6$ and the separation time of two pulses $\mu = \tau / 10$ are used. Fig.~\ref{Figure2}(a) and Fig.~\ref{Figure2}(c) gives the pulse structures of the optimized and conventional STIRAP schemes, respectively.
The optimized STIRAP pulse has the maximum intensity of the Stokes pulse from the beginning of the evolution, while the intensity of the pump pulse is zero. As the evolution goes on, the intensity of the Stokes pulse decreases according to $\cos(\pi s /2)$, and the intensity of the pump pulse increases according to $\sin(\pi s / 2)$. In contrast, the intensities of the Stokes and pump pulses are zeroes for the traditional STIRAP at the beginning of evolution, but the intensity of the Stokes pulse will reach the maximum first, and then the intensity of the pump pulse will reach the maximum, which is so called a counter-intuitive pulse sequence. The population evolutions of the optimized STIRAP and traditional STIRAP are shown in Fig.~\ref{Figure2}(b) and Fig.~\ref{Figure2}(d). It can be seen that the efficiency of the optimized STIRAP to transfer the population to the state $\ket{3}$ is more than 98\%, while the transfer efficiency of the traditional STIRAP is about 72\%. The reason is that the operation time $\tau$ of the system is short, and the traditional STIRAP scheme cannot meet the adiabatic condition well.
The nonadiabatic transition rate in the optimized STIRAP scheme is a small constant, so the nonadiabatic transition rate is not very large even in a short operation time, and the adiabatic transfer can be well realized.
\begin{figure}[h]
\centering
\includegraphics[width=0.3\textwidth]{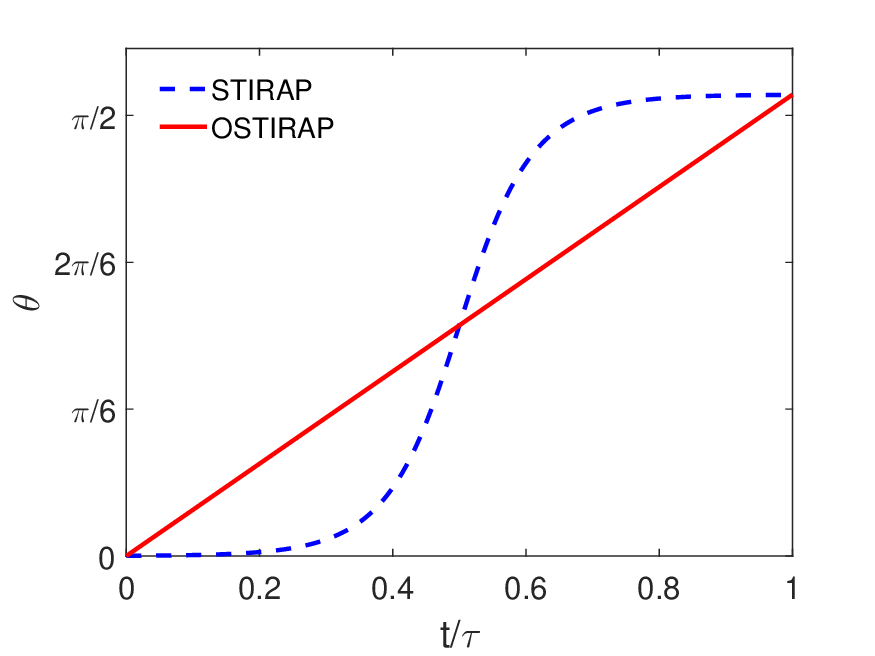}
\caption{Variation of the mixing angle with time (The red solid line indicates the mixing angle for the optimal STIRAP and the blue dashed line indicates the mixing angle for the traditional
 STIRAP)}
\label{Figure3}
\end{figure}

We can further examine the difference between optimized and traditional STIRAP schemes from the change of mixing angle with time. Fig.~\ref{Figure3} gives the change of the mixing angle of the system with time for the two transfer schemes. In the conventional STIRAP scheme, the mixing angle $\theta(t)$ has a large change rate in the $s$ from 0.3 to 0.7, which will destroy the local adiabatic condition, increase the nonadiabatic transition rate of the system, and continuously transfer the population to the bright state channel. The equivalent mixing angle $\alpha s$ in the optimized STIRAP scheme is a straight line passing through the origin with a slope of $\alpha$. At the end of the evolution, there are $\alpha s=\pi/2$.
\begin{figure}[h]
\centering
\includegraphics[width=0.3\textwidth]{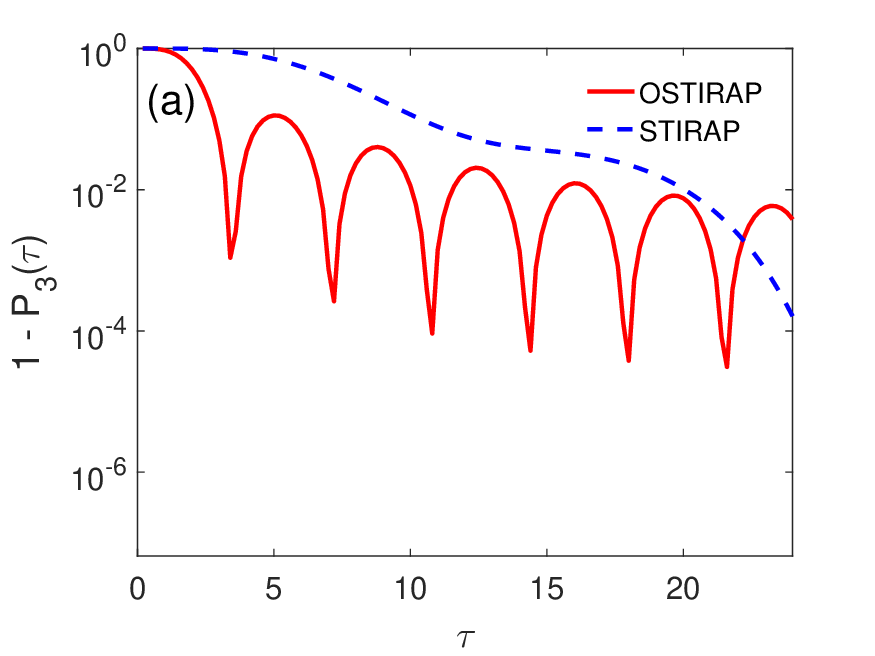}
\includegraphics[width=0.3\textwidth]{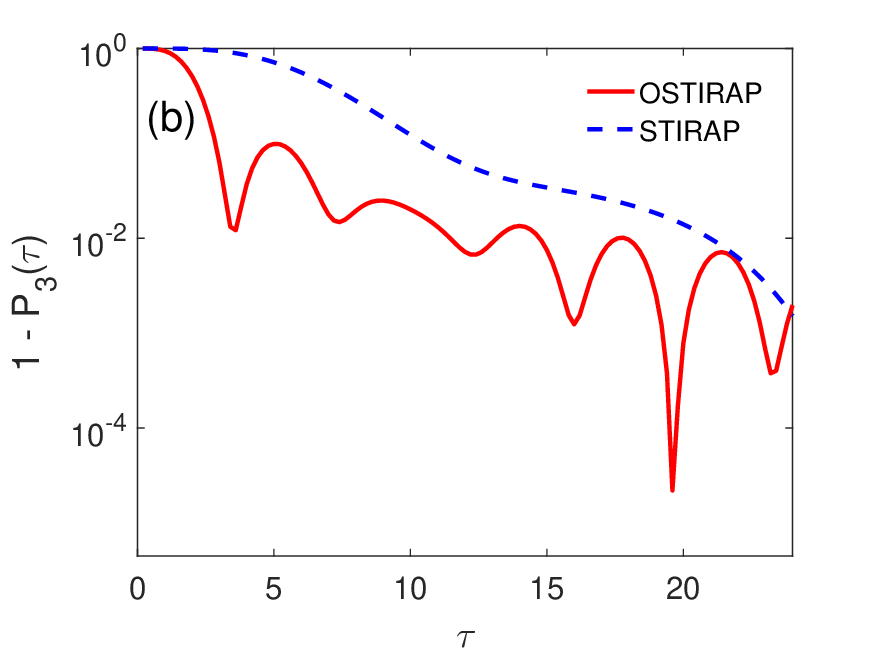}
\caption{Change of the infidelity with time for the three-level system: (a) The case without detuning ($\Delta$=0); (b) The case with detuning ($\Delta$=2$\pi$MHz) (The red solid line corresponds to the optimal STIRAP scheme and the blue dashed line corresponds to the traditional STIRAP one; The pulse peak $\Omega_{0}$=35MHz) }
\label{Figure4}
\end{figure}

\subsection{C. Effect of external field parameters on population transfer process}
Fig.~\ref{Figure4}(a) gives the change of the infidelity of the system with the operation time of the pulse in the absence of detuning. The infidelity is defined as the total population minus the population on the state $\ket{3}$ after the evolution.
The smaller the infidelity, the higher the transfer efficiency, and the more stable the transfer process for the change of parameters.
Fig.~\ref{Figure4} shows that the infidelity of the optimized STIRAP changes periodically with $\tau$. With the extension of the operation time, the highest point of the infidelity decreases gradually, and the lowest point corresponds to a series of time windows. By selecting the operation time within these time windows, the infidelity of the optimized STIRAP can be kept below $10^{-3}$. For example, in the first time window, when $\tau = 3.7\mu\text{s}$ is selected, the infidelity of the optimized STIRAP can be reduced to below $10^{-3}$. However, in the traditional STIRAP scheme with the Gaussian pulses, the overall infidelity decreases slowly, and the infidelity of the transfer process cannot be reduced below $10^{-3}$ until the operation time is $23.8\mu\text{s}$.
\begin{figure}[h]
\centering
\includegraphics[width=0.3\textwidth]{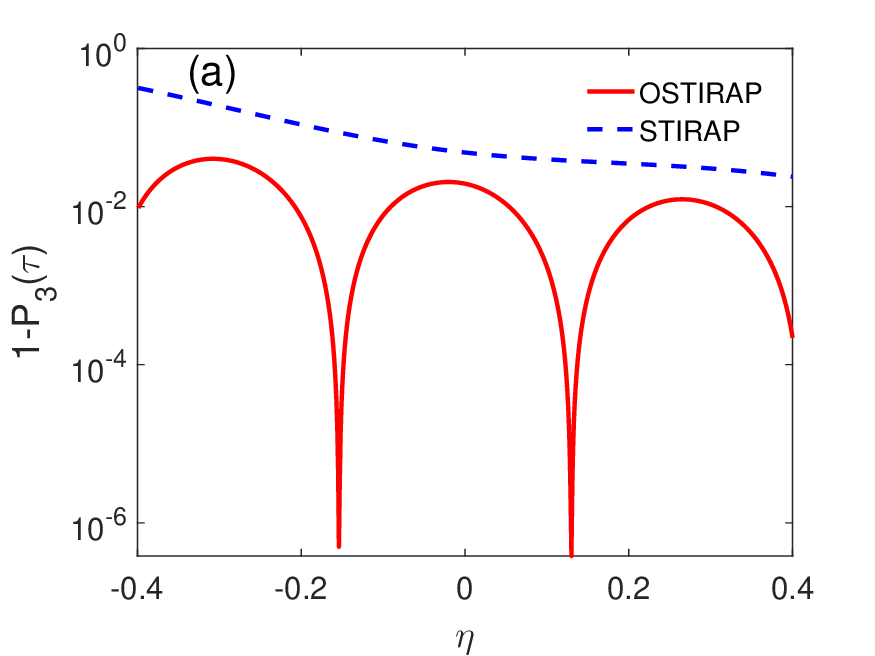}
\includegraphics[width=0.3\textwidth]{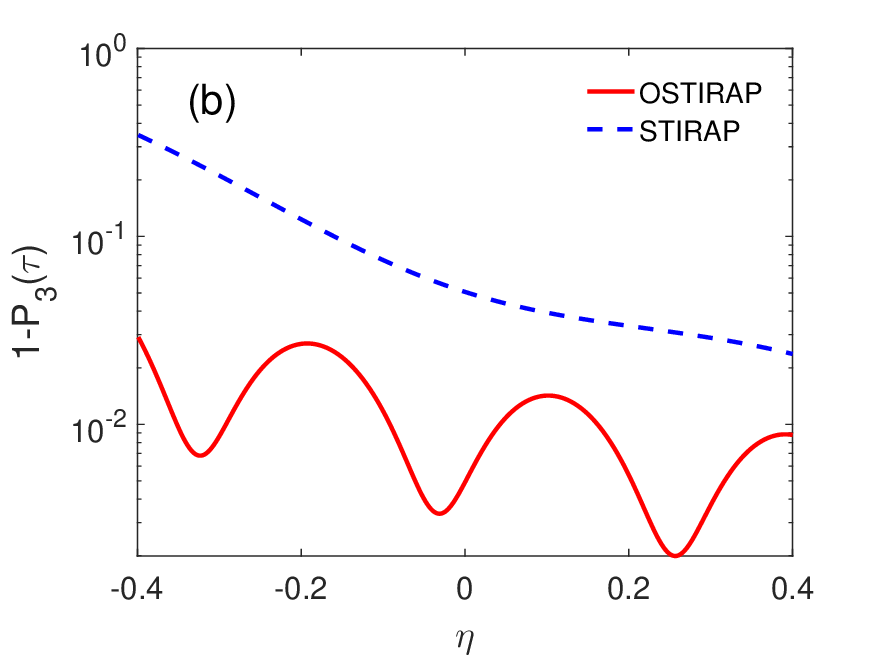}
\caption{Variation of the infidelity with the fluctuation of the pulse peak for the three-level system: (a) The case without detuning ($\Delta$=0); (b) The case with detuning ($\Delta$=2$\pi$MHz) (The red solid line denotes the optimal STIRAP scheme and the blue dashed line denotes the traditional STIRAP one; The pulse peak $\Omega_{ 0 }$=35MHz and the operating time $\tau$=7.4$\mu s$) }
\label{Figure5}
\end{figure}

\begin{figure}[h]
\centering
\includegraphics[width=0.3\textwidth]{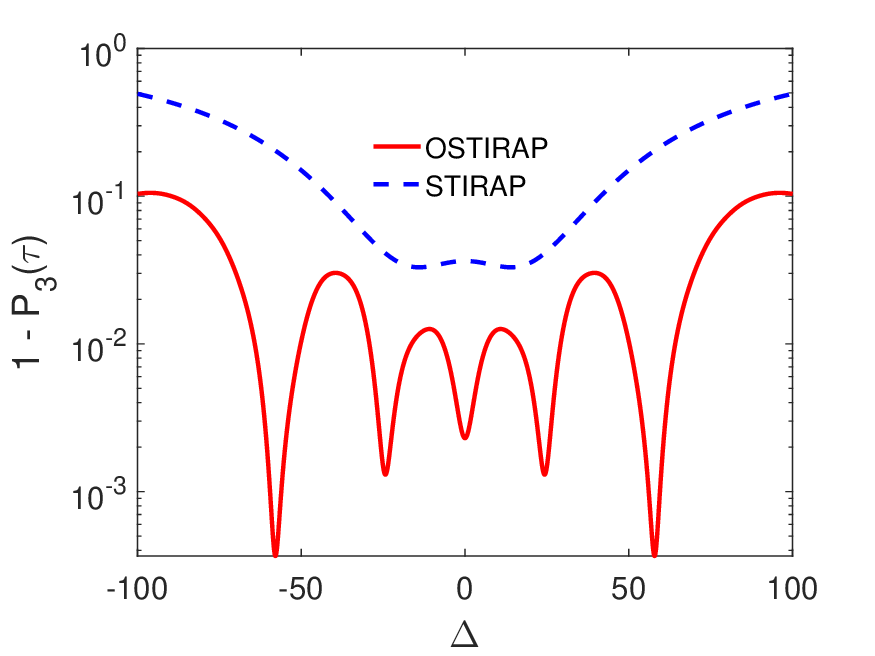}
\caption{Variation of the infidelity with single-photon detuning for the three-level system (The red solid line denotes the optimal STIRAP scheme and the blue dashed line denotes the traditional STIRAP one; The pulse peak $\Omega_0=$35MHz and the pulse operating time $\tau$=7.4$\mu$s)}
\label{Figure6}
\end{figure}
The results show that the optimized STIRAP can achieve nearly complete population transfer in a relatively short time by selecting a specific operating window at the very low point. Fig.~\ref{Figure4}(b) gives the variation of the infidelity with the operation time in the presence of the detuning. When the detuning is not zero, the periodicity of the infidelity for the optimized STIRAP is affected, and the overall infidelity moves up, but there are still some very low points of the infidelity. In the corresponding low operating window, the system can still achieve enough population transfer, so that the infidelity is reduced to below $10^{-2}$. In a word, the optimized STIRAP can perform more complete population transfer in a shorter time, while the existence of detuning will significantly reduce the transfer efficiency of the process.

\subsection{D. Stability of population transfer in a three-level system}
Next, the stability of the optimized STIRAP and traditional STIRAP on the system's parameters is studied.
Firstly, the influence of the fluctuation of the pulse peak $\Omega_0$ on the optimized STIRAP is analyzed.
The peak value of the pulse is taken as $\Omega_0(1+\eta)$. The variation of the infidelity with the fluctuation $\eta$ is given in Fig.~\ref{Figure5}. The selected pulse operation time is $\tau=7.4\mu\text{s}$, which is in the second time window of Fig.~\ref{Figure4}(a). Fig.~\ref{Figure5}(a) gives the change of infidelity with $\eta$ when the detuning is $\Delta = 0$.
The infidelity distribution of the optimized STIRAP is asymmetric about $\eta=0$, and the overall trend decreases slightly with the increase of $\eta$. The infidelity of the traditional STIRAP decreases with the increase of $\eta$. The infidelity of the optimized STIRAP is always less than that of the traditional STIRAP.
At $\eta=-0.15$ and $0.13$, the infidelity of the optimized STIRAP has a very low resonance point and can be reduced to below $10^{-6}$. Fig.~\ref{Figure5}(b) gives the variation of the infidelity with $\eta$ when the detuning is $\Delta=2\pi\text{MHz}$. It can be seen that with the appearance of the detuning, the infidelity of the optimized STIRAP increases as a whole, it's average value is about $10^{-2}$. The resonance peak is obviously broadened, and the infidelity of the optimized STIRAP is still smaller than that of the traditional STIRAP in the presence of the detuning.

The effect of a variation of the detuning $\Delta$ on the optimized STIRAP is considered below. Similarly, the infidelity is used as the criterion for judging the transfer efficiency. From Fig.~\ref{Figure6}, it can be seen that the infidelity of the optimized STIRAP demonstrates a series of resonance windows, and the infidelities of all the resonance points can be reduced to lower than $10^{-2}$. In the traditional STIRAP, the  infidelity increases with increasing the absolute value of the detuning. The transition trends in both cases are symmetric about $\Delta=0$. For any value of the detuning, the infidelity of the traditional STIRAP process is greater than that of the optimized STIRAP process.
\\
\\
\\
\section{IV. Optimized STIRAP scheme for a four-level system}
\subsection{A. Decoupling method for degenerate double dark states of a four-level system}
\begin{figure}[h]
\centering
\includegraphics[width=0.4\textwidth]{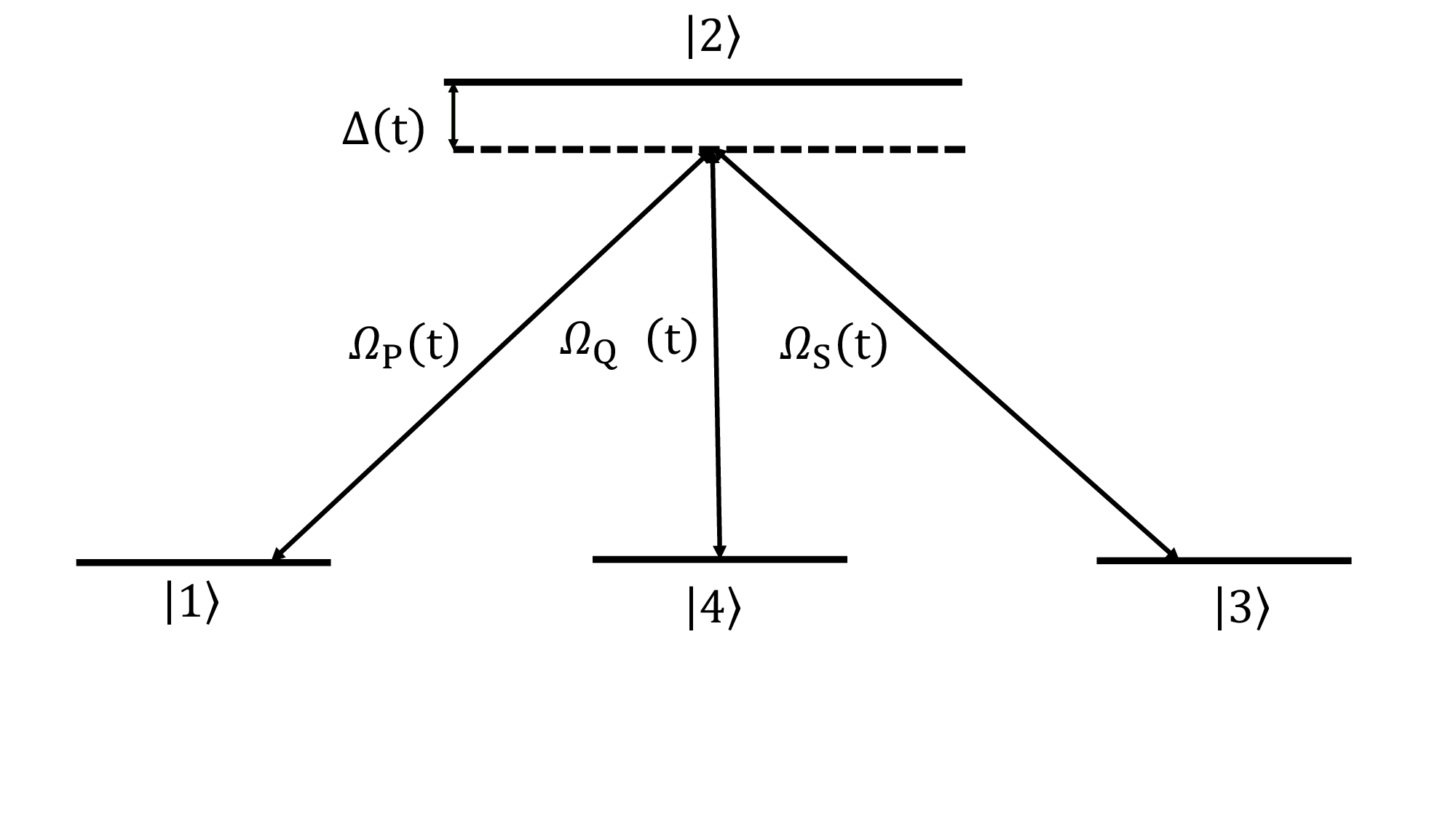}
\caption{Schematic diagram of the STIRAP scheme for a four-level system with single photon detuning}
\label{Figure7}
\end{figure}
Consider the adiabatic population transfer process of a tripod-type four-level system indicated by Fig.~\ref{Figure7}.
The system can realize population transfer from the state $\ket{1}$ to the states $\ket{3}$ and $\ket{4}$, and can also form quantum superposition states of the states $\ket{3}$ and $\ket{4}$ in any proportion.
In the rotating-wave approximation, the Hamiltonian of this four-level system is~\cite{BWS2013,KBer2015,CSMad2024,ZCShi2024,ZYJin2025,GQLi2025}
\begin{equation}
	H(t)=
	\frac{\hbar}{2}
	\begin{pmatrix}
		0            & \Omega_{P}(t)     & 0               & 0 \\
		\Omega_{P}^{*}(t) & \Delta(t)     & \Omega_{S}(t)   & \Omega_{Q}(t) \\
		0            & \Omega_{S}^{*}(t) & 0               & 0 \\
		0            & \Omega_{Q}^{*}(t) & 0               & 0
	\end{pmatrix},\label{Formula12}
\end{equation}
where $\Omega_P(t)$, $\Omega_S(t)$ and $\Omega_Q(t)$ are Rabi pulses between the coupling ground states $\ket{1}$, $\ket{3}$ and $\ket{4}$ and the excited state $\ket{2}$. Compared with the three-level system, the adiabatic transfer for the four-level system is slightly more complicated due to the existence of two degenerate dark states.
However, under certain conditions, the four-level system can be reduced into a three-level one.
In this paper, a method for eliminating the influence of degenerate energy levels for a four-level system is given~\cite{NVV2017}.
The method is also applicable to systems with more energy levels under the same circumstance.

For the above four-level Hamiltonian, its four eigenvalues are
$\lambda_{+} = -\lambda_{-} = \Omega / 2$ and $\lambda_{1} = \lambda_{2} = 0$,
where $\Omega = \sqrt{\Omega_P^2 + \Omega_S^2 + \Omega_Q^2}$.
The corresponding eigenstates are
\begin{widetext}
	\begin{align}
		\ket{\lambda_{+}} &=
		\sin\vartheta(t)\sin\phi(t)\ket{1}
		+ \cos\phi(t)\ket{2}
		+ \sin\phi(t)\cos\vartheta(t)\cos\xi(t)\ket{3}
		+ \sin\phi(t)\cos\vartheta(t)\sin\xi(t)\ket{4}, \nonumber\\[4pt]
		\ket{\lambda_{1}} &=
		\cos\vartheta(t)\ket{1}
		- \sin\vartheta(t)\cos\xi(t)\ket{3}
		- \sin\vartheta(t)\sin\xi(t)\ket{4}, \nonumber\\[4pt]
		\ket{\lambda_{2}} &=
		\sin\xi(t)\ket{3}
		- \cos\xi(t)\ket{4}, \nonumber\\[4pt]
		\ket{\lambda_{-}} &=
		\cos\phi(t)\sin\vartheta(t)\ket{1}
		- \sin\phi(t)\ket{2}
		+ \cos\phi(t)\cos\vartheta(t)\cos\xi(t)\ket{3}
		+ \cos\phi(t)\cos\vartheta(t)\sin\xi(t)\ket{4}, \label{Formula13}
	\end{align}
\end{widetext}
where the mixing angles satisfy
$\tan\xi(t)= \Omega_{\mathrm{Q}}(t) / \Omega_{\mathrm{S}}(t)$,
$\tan\vartheta(t) = \Omega_{\mathrm{P}}(t) / \sqrt{\Omega_{\mathrm{S}}^2(t) + \Omega_{\mathrm{Q}}^2(t)}$,
and $\tan2\phi(t)= \Omega(t) / \Delta(t)$.

In this system, the eigenstates $\ket{\lambda_{1,2}}$ corresponding to the two zero eigenvalues $\lambda_{1,2}$ are called the dark states, and the eigenstates $\ket{\lambda_{\pm}}$ corresponding to the two nonzero eigenvalues $\lambda_{\pm}$ are called the bright states. Since the two eigenstates $\ket{\lambda_{1,2}}$ with zero eigenvalues are degenerate,
the nonadiabatic coupling between them cannot be suppressed even if the adiabatic condition is satisfied~\cite{NVV2017}.
The coupling between the eigenstates $\ket{\lambda_1}$ and $\ket{\lambda_2}$ will seriously affect the efficiency of the final population transfer, so it is necessary to find a way to eliminate this coupling.

In order to establish the relationship between the degenerate energy levels, we transform the system from the pure state space to the eigenstate space. The unitary transformation matrix is~\cite{BWS1991}
\begin{widetext}
	\begin{equation}
		U(t) =
		\begin{pmatrix}
			\cos\vartheta & 0 & \sin\phi\sin\vartheta & \cos\phi\sin\vartheta \\[4pt]
			0 & \cos\phi & -\sin\phi & 0 \\[4pt]
			-\sin\vartheta\cos\xi & \sin\xi & \sin\phi\cos\vartheta\cos\xi & \cos\phi\cos\vartheta\cos\xi \\[4pt]
			-\sin\vartheta\sin\xi & -\cos\xi & \sin\phi\cos\vartheta\sin\xi & \cos\phi\cos\vartheta\sin\xi
		\end{pmatrix}.\label{Formula14}
	\end{equation}
\end{widetext}
The transformation relation of the system's Hamiltonian is
$\tilde{H}(t) = {U(t)}^{-1} H(t) U(t) + i \dot{U}^{-1}(t) U(t)$,
which can be obtained by substituting~(\ref{Formula13}) in it. The transformed Hamiltonian is expressed as
\begin{widetext}
	\begin{equation}
		\tilde{H}(t) = -i
		\begin{pmatrix}
			0 & \dot{\xi}\sin\vartheta & \dot{\vartheta}\cos\xi & \dot{\vartheta}\sin\xi \\[4pt]
			-\dot{\xi}\sin\vartheta & 0 & -\dot{\phi}\cos\vartheta\cos\xi & -\dot{\phi}\cos\vartheta\sin\xi \\[4pt]
			-\dot{\vartheta}\cos\xi & \dot{\phi}\cos\vartheta\cos\xi & -\tfrac{i}{2}\dot{\Omega} & -\dot{\xi} \\[4pt]
			-\dot{\vartheta}\sin\xi & \dot{\phi}\cos\vartheta\sin\xi & \dot{\xi} & \tfrac{i}{2}\dot{\Omega}
		\end{pmatrix}.
	\end{equation}
\end{widetext}
The Schr\"{o}dinger equation satisfied by the eigenstates is
\[
\frac{d}{dt}\,\vec{\lambda}(t)
= -i\,\widetilde{H}(t)\,\vec{\lambda}(t),
\]
where $\vec{\lambda}(t) = [\,\ket{\lambda_1(t)},\,\ket{\lambda_2(t)},\,\ket{\lambda_+(t)},\,\ket{\lambda_-(t)}\,]^{T}$.
In this case, if we only need to study the coupling between $\ket{\lambda_1(t)}$ and $\ket{\lambda_2(t)}$, the reduced equations are obtained:
\begin{equation}
	\frac{\mathrm{d}}{\mathrm{d}t}
	\begin{pmatrix}
		|\lambda_1(t)\rangle \\[2pt]
		|\lambda_2(t)\rangle
	\end{pmatrix}
	=
	\begin{pmatrix}
		0 & -\dot{\xi}(t)\sin\vartheta(t) \\[2pt]
		\dot{\xi}(t)\sin\vartheta(t) & 0
	\end{pmatrix}
	\begin{pmatrix}
		|\lambda_1(t)\rangle \\[3pt]
		|\lambda_2(t)\rangle
	\end{pmatrix}.
\end{equation}
For the reduced Schr\"{o}dinger equation, the general solutions are:
\begin{equation}
	\begin{aligned}
		|\lambda_1(t)\rangle &= |\lambda_1(-\infty)\rangle \cos\Theta(t)
		- |\lambda_2(-\infty)\rangle \sin\Theta(t),\\
		|\lambda_2(t)\rangle &= |\lambda_2(-\infty)\rangle \cos\Theta(t)
		- |\lambda_1(-\infty)\rangle \sin\Theta(t),
	\end{aligned}
\end{equation}
where $\Theta(t)=\int_{-\infty}^{t}dt^{\prime}\dot{\xi}(t^{\prime})\sin\vartheta(t).$

We have obtained a coupling relation between the two degenerate eigenstates, in which the new mixing angle $\Theta(t)$ is the key involving the change of the two degenerate eigenstates. If $\Theta(t)=0$, the coupling between the two degenerate states disappears, that is, $\ket{\lambda_{1,2}(t)} = \ket{\lambda_1(-\infty)}$, which will satisfy the requirement of adiabatic transfer.
It can be seen from Eq.~(\ref{Formula13}) that there are only the states $\ket{3}$ and $\ket{4}$ but the state $\ket{1}$ doesn't present in the dark state $\ket{\lambda_2(t)}$, so it should be ensured that the adiabatic transfer always evolves along the dark state $\ket{\lambda_1(t)}$, and the population initially populated on the state $\ket{1}$ is efficiently transferred to the states $\ket{3}$ and $\ket{4}$ excluding the influence of $\ket{\lambda_2(t)}$.  As long as $\dot{\xi}(t) = 0$ is guaranteed, the nonadiabatic coupling between the two degenerate states is released due to $\Theta(t)=0$,
and the evolution can proceed along the selected eigenstate $\ket{\lambda_1(t)}$. So the condition for eliminating the nonadiabatic coupling between two degenerate states is
\begin{equation}
\tan\xi(t)=\frac{\Omega_Q(t)}{\Omega_S(t)} = \text{Constant}. \label{Formula18}
\end{equation}

\subsection{B. Quantum geometric tensor of four-level system and calculation of nonadiabatic transition rate}
We have given the condition to eliminate the nonadiabatic transition between the degenerate dark states, then the optimization of the dynamical quantum geometry tensor will be carried out, and the elimination condition of the eigenstate degeneracy~(\ref{Formula18}) will be satisfied at the end. According to~(\ref{DQGT}), ~(\ref{Formula12}), and~(\ref{Formula13}), the components of the dynamical quantum geometric tensor of the dark state $\ket{\lambda_1(t)}$ can be obtained~\cite{KZLi2024,JFChen2022}:
\begin{equation}
	\begin{aligned}
		D_{1,PP} &=
		\frac{4}{\hbar^{2}}
		\frac{\Omega_S^{2} + \Omega_Q^{2}}
		{\Omega_P^{2} + \Omega_S^{2} + \Omega_Q^{2}}
		\frac{M_{+} + M_{-}}{M_{+}M_{-}}, \\[4pt]
		D_{1,SS} &=
		\frac{4}{\hbar^{2}}
		\frac{\Omega_P^{2}\Omega_S^{2}}
		{(\Omega_P^{2} + \Omega_S^{2} + \Omega_Q^{2})(\Omega_S^{2} + \Omega_Q^{2})}
		\frac{M_{+} + M_{-}}{M_{+}M_{-}}, \\[4pt]
		D_{1,QQ} &=
		\frac{4}{\hbar^{2}}
		\frac{\Omega_P^{2}\Omega_Q^{2}}
		{(\Omega_P^{2} + \Omega_S^{2} + \Omega_Q^{2})(\Omega_S^{2} + \Omega_Q^{2})}
		\frac{M_{+} + M_{-}}{M_{+}M_{-}}, \\[4pt]
		D_{1,PS} &= D_{1,SP} =
		-\frac{4}{\hbar^{2}}
		\frac{\Omega_P \Omega_S}
		{\Omega_P^{2} + \Omega_S^{2} + \Omega_Q^{2}}
		\frac{M_{+} + M_{-}}{M_{+}M_{-}}, \\[4pt]
		D_{1,PQ} &= D_{1,QP} =
		-\frac{4}{\hbar^{2}}
		\frac{\Omega_P \Omega_Q}
		{\Omega_P^{2} + \Omega_S^{2} + \Omega_Q^{2}}
		\frac{M_{+} + M_{-}}{M_{+}M_{-}}, \\[4pt]
		D_{1,SQ} &= D_{1,QS} =
		\frac{4}{\hbar^{2}}
		\frac{\Omega_P^{2}\Omega_S \Omega_Q}
		{(\Omega_P^{2} + \Omega_S^{2} + \Omega_Q^{2})(\Omega_S^{2} + \Omega_Q^{2})}
		\frac{M_{+} + M_{-}}{M_{+}M_{-}} .
	\end{aligned}
\end{equation}
From Eq.~(\ref{Formula4}), we can obtain the total nonadiabatic transition rate
\[		\tilde{T}_{n_1}^{2}(s)
		= \frac{4}{\hbar^{2}\Omega^{2}}
		\frac{M_{+}+M_{-}}{M_{+}M_{-}}
		\Bigl[
		(\Omega_{S}^{2}+\Omega_{Q}^{2})\dot{\Omega}_{P}^{2}  \]
\begin{equation}
+ \frac{\Omega_{P}^{2}}{\Omega_{S}^{2}+\Omega_{Q}^{2}}(\Omega_{S}\dot{\Omega}_{S}+\Omega_{Q}\dot{\Omega}_{Q})^{2}
		- 2\Omega_{P}\dot{\Omega}_{P}(\Omega_{S}\dot{\Omega}_{S}+\Omega_{Q}\dot{\Omega}_{Q})
		\Bigr].
\end{equation}
When the system detuning $\Delta = 0$, the total nonadiabatic transition rate will be transformed into
\[		\tilde{T}_{n_1}^{2}(s)
		= \frac{4}{\hbar^{2}\Omega^{2}}
	     \Bigl[
		(\Omega_{S}^{2}+\Omega_{Q}^{2})\dot{\Omega}_{P}^{2}    \]
\[		+ \frac{\Omega_{P}^{2}}{\Omega_{S}^{2}+\Omega_{Q}^{2}}(\Omega_{S}\dot{\Omega}_{S}+\Omega_{Q}\dot{\Omega}_{Q})^{2}
		- 2\Omega_{P}\dot{\Omega}_{P}(\Omega_{S}\dot{\Omega}_{S}+\Omega_{Q}\dot{\Omega}_{Q})
		\Bigr].   \]
To optimize the total nonadiabatic transition rate, two conditions need to be satisfied, namely, (\ref{Formula18}) and $\tilde{T}_{n1}^2(s)$ is constant.
Pulses satisfying these two conditions can realize adiabatic evolution along the dark state $\ket{\lambda_1(t)}$, and the influence of nonadiabatic transitions will be minimized.
In this paper, a combination of Rabi pulses satisfying this condition is given for a four-level system:
\begin{equation}
	\begin{aligned}
		\tilde{\Omega}_{P}(t) &= \Omega_{0}\sin(\beta s),\\
		\tilde{\Omega}_{S}(t) &= \Omega_{0}\cos(\beta s)\cos\chi,\\
		\tilde{\Omega}_{Q}(t) &= \Omega_{0}\cos(\beta s)\sin\chi.
	\end{aligned}
\end{equation}
Using the above combination of pulses, the total nonadiabatic transition rate becomes
		\[\tilde{T}_{n_1}^{2}(s)
		= \frac{2\Omega_{0}}{\hbar^{2}}
		\Bigl(
		\Bigl\{
		[\Omega_{0}^{2}+(\Delta+\sqrt{\Delta^{2}+\Omega_{0}^{2}})^{2}]
		[(\Delta+\sqrt{\Delta^{2}+\Omega_{0}^{2}})^{2}]\Bigr\}^{-1}\]
	\[	+
		\Bigl\{
		[\Omega_{0}^{2}+(\Delta-\sqrt{\Delta^{2}+\Omega_{0}^{2}})^{2}]
		[(\Delta-\sqrt{\Delta^{2}+\Omega_{0}^{2}})^{2}]
		\Bigr\}^{-1}
		\Bigr)^{1/2}.\]
The parameter $\chi$ in the optimized pulse is a controllable parameter, which is equivalent to the mixing angle $\xi$ in the traditional STIRAP process. Taking another parameter $\alpha=\pi/2$, the evolution of the system will proceed along the dark state $\ket{\lambda_1(t)}$, and eventually the population in the initial state $\ket{1}$ will be transformed to the states $\ket{3}$ and $\ket{4}$. From (\ref{Formula13}), we can see that after the evolution, the population on the state $\ket{3}$ is $\cos^2 \chi$, and the population on the state $\ket{4}$ is $\sin^2 \chi$. Therefore, the quantum superposition of the states $\ket{3}$ and $\ket{4}$ in any proportion can be realized by controlling the pulse parameter $\chi$
to design the population ratio of the states $\ket{3}$ and $\ket{4}$ after the evolution.

\subsection{C. Population transfer dynamics of a four-level system}
\begin{figure}[h]
\centering
\includegraphics[width=0.2\textwidth]{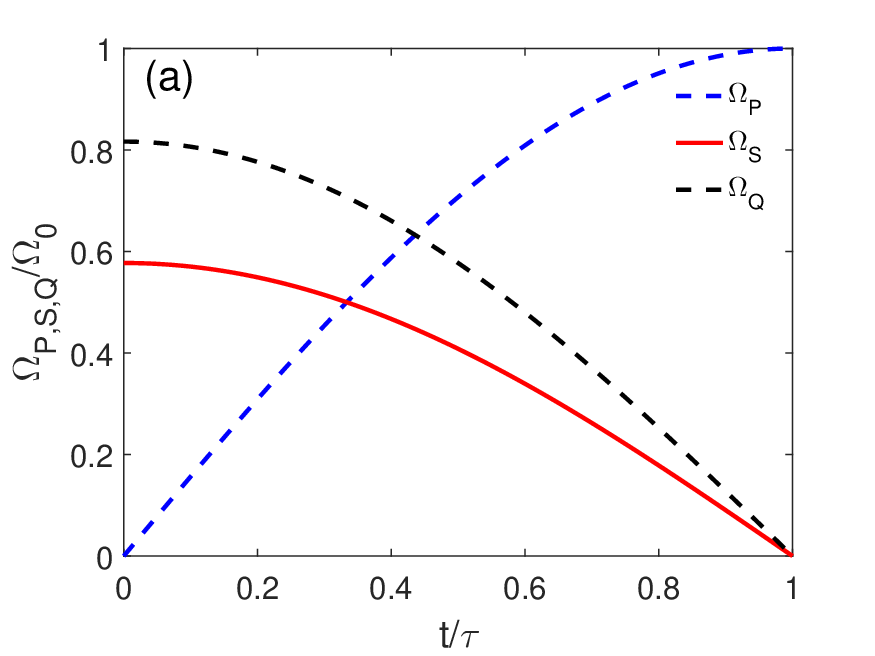}
\includegraphics[width=0.2\textwidth]{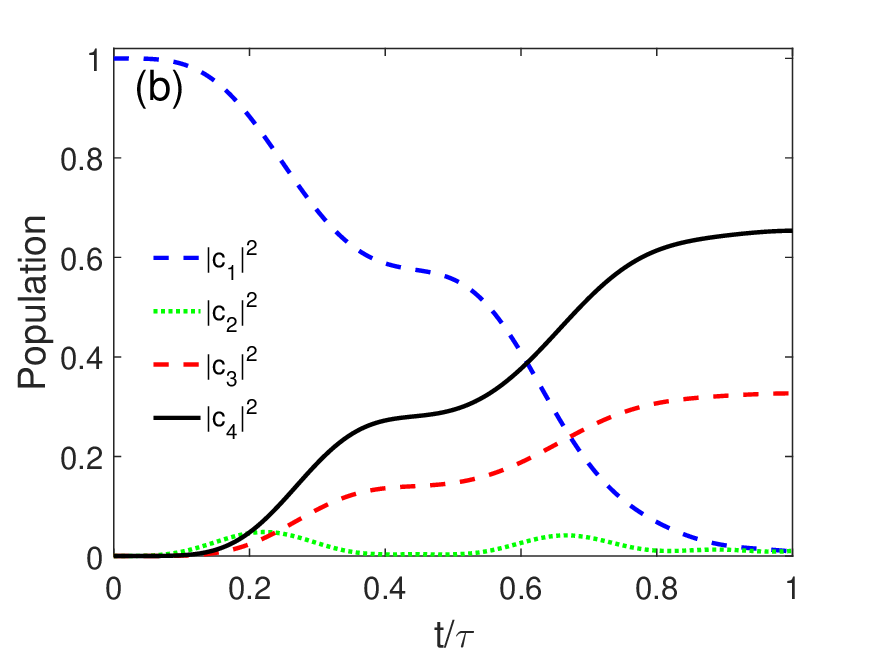}
\includegraphics[width=0.2\textwidth]{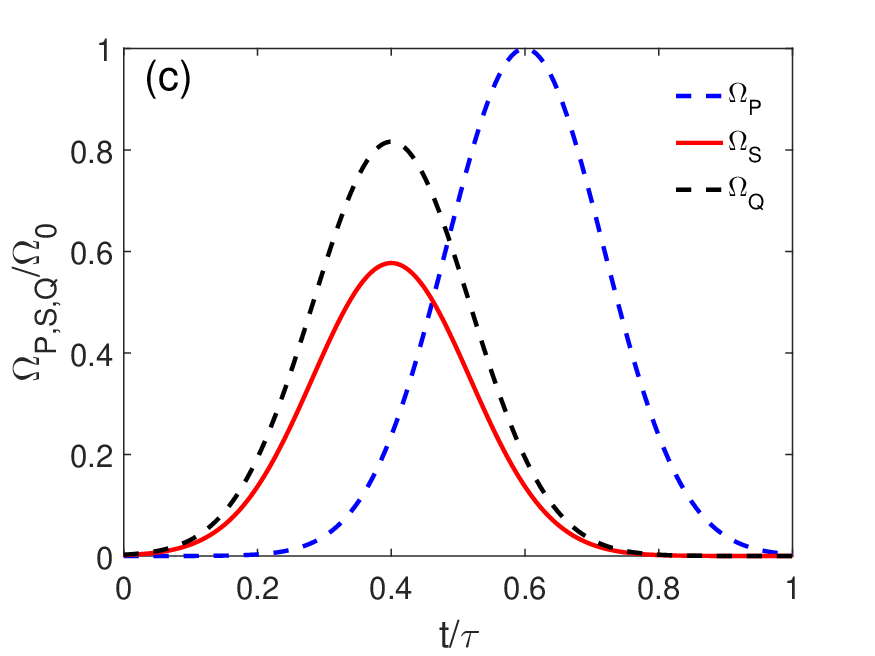}
\includegraphics[width=0.2\textwidth]{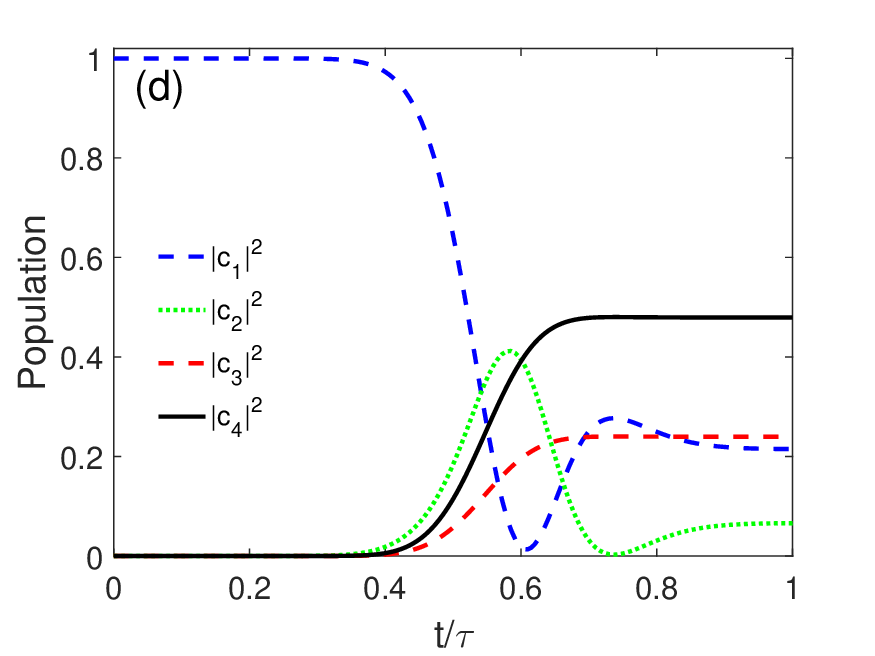}
\caption{Rabi pulse's structures and the evolution results of the populations for the four-level system: (a) Pulse structures for the optimal STIRAP; (b) Evolution of the populations for the optimal STIRAP; (c) Pulse structures for the traditional STIRAP; (d) Evolution of the populations for the traditional STIRAP (The pulse operating time $\tau=4\mu s$, the pulse peak $\Omega_0=$35MHz, and the detuning $\Delta$=2$\pi$MHz) }
\label{Figure8}
\end{figure}

It is assumed that all the particles in the initial state are populated on the state $\ket{1}$.
According to the pulse sequence given above, if the parameter $\chi = \arccos(1/3)$ is selected,
the population on the final state $\ket{3}$ is $1/3$, and the population on the state $\ket{4}$ is $2/3$.
The Gaussian pulses used in the traditional STIRAP process is given as follows:
\begin{equation}
	\begin{aligned}
		\Omega_{P}(t) &= \Omega_{0}\exp[-(t-\tau/2-\mu)^{2}/\sigma^{2}],\\
		\Omega_{S}(t) &= \Omega_{0}\exp[-(t-\tau/2+\mu)^{2}/\sigma^{2}]\cos\chi,\\
		\Omega_{Q}(t) &= \Omega_{0}\exp[-(t-\tau/2+\mu)^{2}/\sigma^{2}]\sin\chi.\label{Formula22}
	\end{aligned}
\end{equation}
Similar to the three-level system, the full width at half maximum of the Gaussian pulses $\sigma = \tau / 6$ and the separation time of the two pulses $\mu = \tau / 10$ are taken in the calculation. The pulse structures of the optimized STIRAP and traditional STIRAP are shown in Fig.~\ref{Figure8}(a) and Fig.~\ref{Figure8}(c), respectively. In the optimized STIRAP pulse, the evolution starts when the values of $\Omega_{\mathrm{Q}}$ and $\Omega_{\mathrm{S}}$ reach their maximum values and the value of $\Omega_{\mathrm{P}}$ is zero. With the evolution, $\Omega_{\mathrm{Q}}$ and $\Omega_{\mathrm{S}}$ decrease, while $\Omega_{\mathrm{P}}$ increases. In contrast, the traditional Gaussian pulses has zero values of $\Omega_{\mathrm{P}}$, $\Omega_{\mathrm{Q}}$, and $\Omega_{\mathrm{S}}$ at the beginning of the evolution,
then $\Omega_{\mathrm{Q}}$ and $\Omega_{\mathrm{S}}$ appear simultaneously and keep $\Omega_{\mathrm{Q}}(t) / \Omega_{\mathrm{S}}(t)$ constant, then the role of $\Omega_{\mathrm{P}}$ appears. Fig.~\ref{Figure8}(b) and Fig.~\ref{Figure8}(d) give the evolution process of the population of the optimized STIRAP and traditional STIRAP, respectively.
In the optimized STIRAP evolution, small population fluctuations appear in the intermediate state $\ket{2}$, but these fluctuations quickly disappear, eventually transferring almost all the population to the states $\ket{3}$ and $\ket{4}$.
Because the mixing angle $\xi$ is a constant, the nonadiabatic transition between the degenerate dark states is suppressed,
resulting in a final population of $1/3$ in the state $\ket{3}$ and $2/3$ in the state $\ket{4}$.
For the traditional STIRAP process, a larger population appears in the intermediate state $\ket{2}$, indicating that the system has a larger nonadiabatic transition, so that the evolution does not completely follow the dark state $\ket{\lambda_1}$.
The reason is that the operation time $\tau$ is small, and the traditional STIRAP cannot satisfy the adiabatic condition, which leads to a large nonadiabatic transition, while the optimized STIRAP scheme makes the nonadiabatic transition rate a small constant by means of the optimization of the dynamical quantum geometric tensor, so it has little effect on the evolution results.
\begin{figure}[h]
\centering
\includegraphics[width=0.3\textwidth]{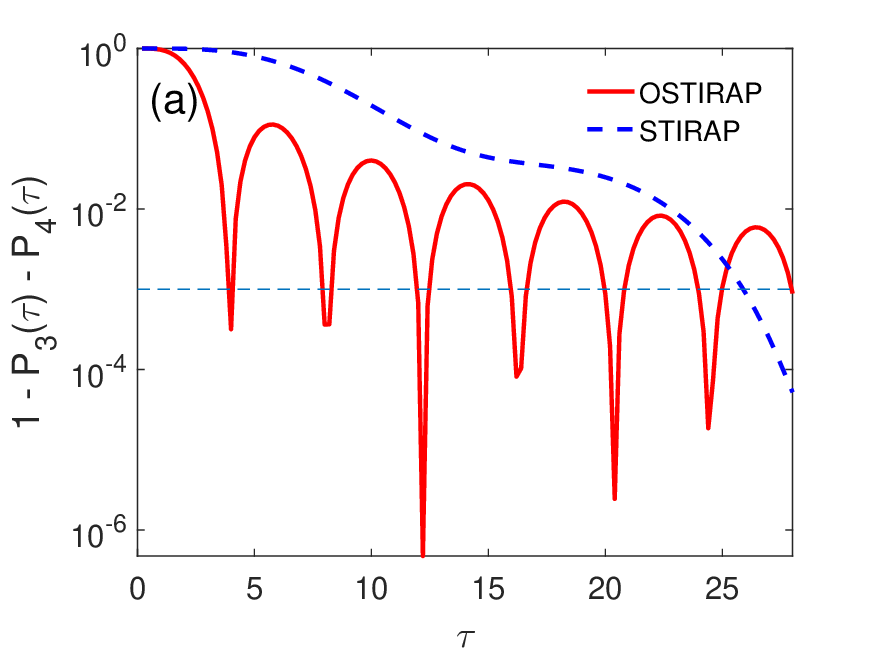}
\includegraphics[width=0.3\textwidth]{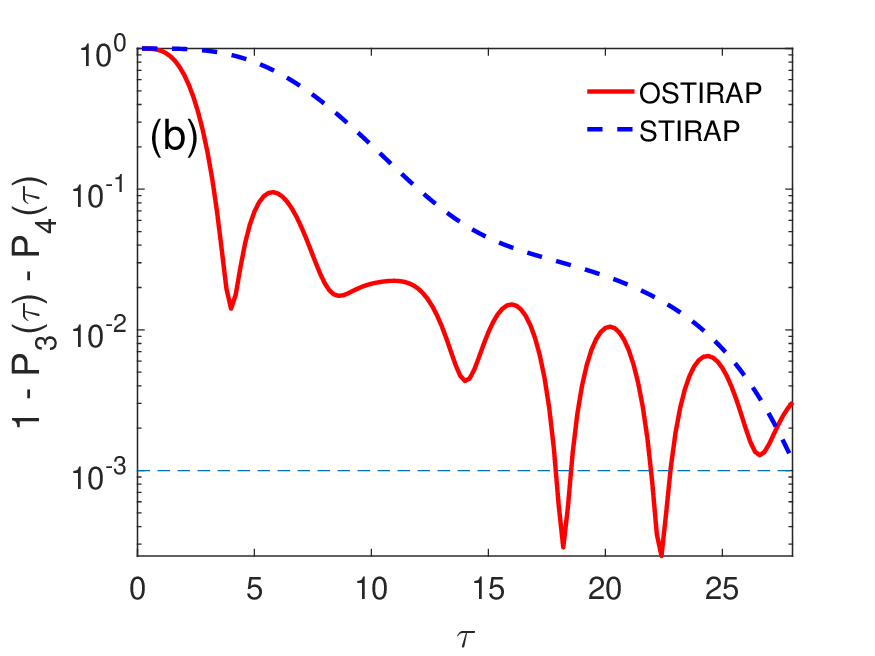}
\caption{Variation of the infidelity with time for the four-level system: (a) The case without detuning ($\Delta$=0); (b) The case with detuning ($\Delta$=2$\pi$MHz) (The red solid line corresponds to the optimal STIRAP scheme and the blue dashed line corresponds to the traditional STIRAP one; The pulse peak $\Omega_0$=35MHz) }
\label{Figure9}
\end{figure}

Fig.~\ref{Figure9}(a) gives the variation of the infidelity with the operation time when the detuning $\Delta=0$.
The infidelity here is defined as one minus the sum of the population on the states $\ket{3}$ and $\ket{4}$ after the evolution.
The smaller the infidelity, the more complete the population transfer of the system. According to Fig.~\ref{Figure9}(a), the infidelity of the optimized STIRAP changes with the operation time periodically. The highest point of the infidelity gradually moves down, and the lowest point corresponds to a series of time windows. The operation times at these time windows are selected such that the infidelity of the optimized STIRAP is kept below $10^{-3}$. For example, in the first time window, the infidelity of the optimized STIRAP is reduced to below $10^{-3}$ after selecting $\tau = 4\mu\text{s}$. However, the traditional STIRAP scheme using the Gaussian pulses shows a slow downward trend, and the infidelity cannot be reduced below $10^{-3}$ until the operation time reaches $25.8\mu\text{s}$. The results show that the optimized STIRAP scheme requires less time for the complete transfer than the traditional STIRAP scheme and can achieve fast and efficient population transfer.
Fig.~\ref{Figure9}(b) gives the variation of infidelity with the operation time when the detuning $\Delta=2\pi\text{MHz}$.
Once the detuning is not zero, the periodicity of the infidelity of the optimized STIRAP scheme is affected, and the infidelities of both the optimized STIRAP and traditional STIRAP are shifted upward. However, there is still a time window for the optimized STIRAP to achieve lower infidelity. In the corresponding very low working window, the infidelity is low, and a more complete population transfer can still be achieved.

\subsection{D. Stability of population transfer in a four-level system}
The optimized STIRAP scheme of the four-level system can realize the superposition of the final states $\ket{3}$ and $\ket{4}$ in an arbitrary proportion, and the transfer time is significantly shorter than that of the traditional STIRAP scheme. Next, the stability of the optimized STIRAP scheme with respect to the fluctuations of the two system parameters will be discussed.
\begin{figure}[h]
\centering
\includegraphics[width=0.3\textwidth]{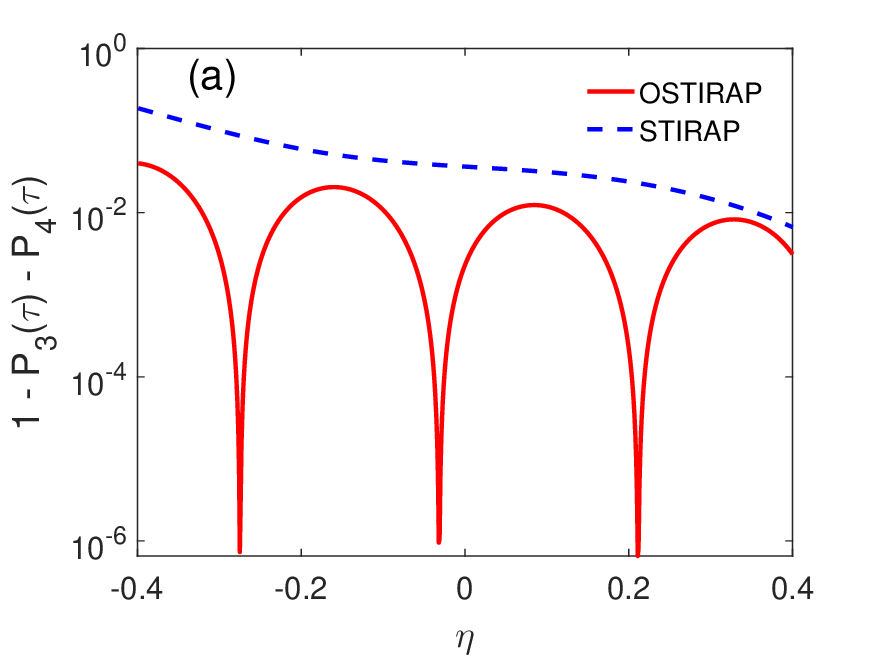}
\includegraphics[width=0.3\textwidth]{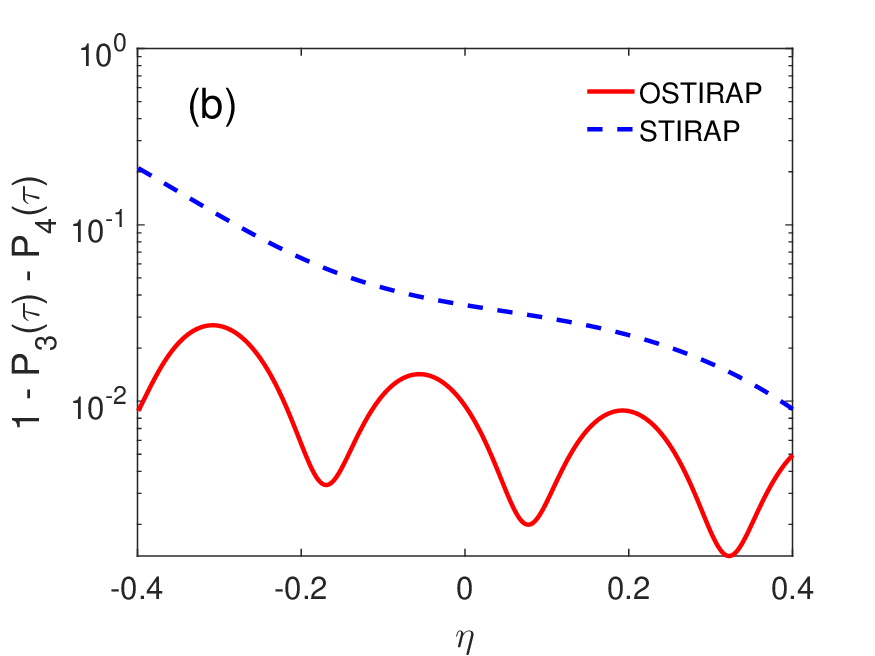}
\caption{Variation of the infidelity with the fluctuation of the pulse peak for the four-level system: (a) The case without detuning ($\Delta$=0); (b) The case with detuning ($\Delta$=2$\pi$MHz) (The red solid line denotes the optimal STIRAP scheme and the blue dashed line denotes the standard STIRAP one; The pulse peak $\Omega_0$=35MHz and the operating time $\tau$=7.4$\mu$s)}
\label{Figure10}
\end{figure}

Fig.~\ref{Figure10}(a) gives the variation of the infidelity with the fluctuation $\eta$ of the pulse peak when the detuning $\Delta=0$. When the fluctuation $\eta$ is not zero, the actual value of the pulse peak becomes $\Omega_0(1+\eta)$.
From Fig.~\ref{Figure10}, it can be seen that the optimized STIRAP scheme still has periodicity for the fluctuation of the pulse peak, but the overall infidelity decreases with the increase of the pulse fluctuation. In contrast, the infidelity in the traditional STIRAP scheme decreases with the increase of fluctuation and does not have the characteristics of periodic change.
The infidelity of the traditional STIRAP scheme is generally greater than that of the optimized STIRAP scheme. This periodic feature of the optimized STIRAP scheme indicates that a more complete population transfer can be achieved using a smaller pulse peak and operation time.

Fig.~\ref{Figure10}(b) gives the variation of infidelity with $\eta$ when the detuning $\Delta=2\pi\text{MHz}$.
It can be seen that with the appearance of the detuning $\Delta$, the infidelities of both the optimized STIRAP and traditional STIRAP move up as a whole. The overall trend of the optimized STIRAP remains unchanged, and the infidelity is still periodic and decreases with the increase of the intensity of the pulse. The infidelity of the traditional STIRAP decreases with the increase of pulse peak value, but it is still larger than that of the optimized STIRAP. This shows that the population transfer ability of the optimized STIRAP scheme decreases with the increase of the detuning, but it can still achieve more complete population transfer under short operation time and low pulse peak value by selecting appropriate parameters.

\begin{figure}[h]
\centering
\includegraphics[width=0.3\textwidth]{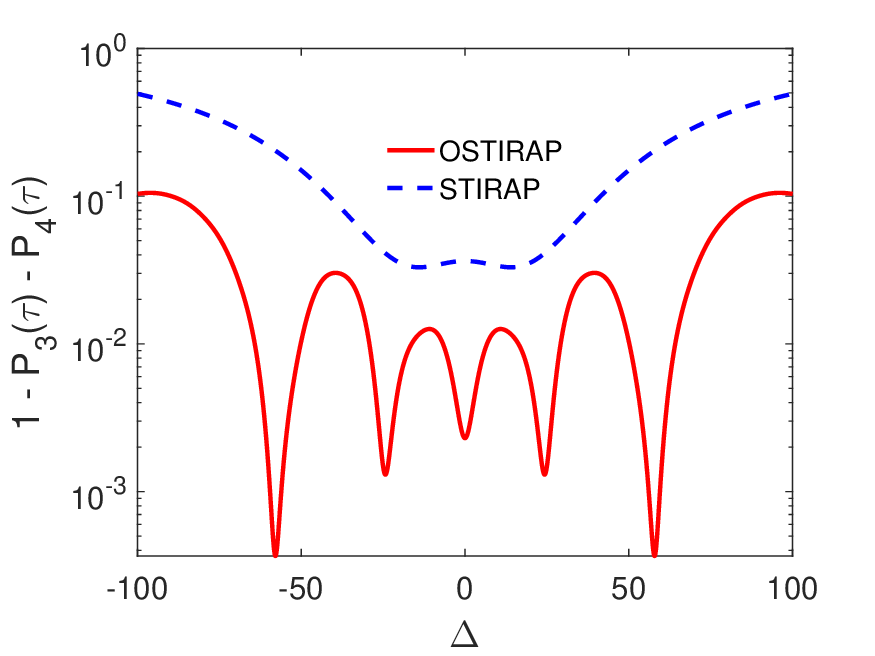}
\caption{Variation of the infidelity with single-photon detuning for the four-level system (The red solid line denotes the optimal STIRAP scheme and the blue dashed line denotes the traditional STIRAP one; The pulse peak $\Omega_0=$35MHz and the pulse operating time $\tau$=7.4$\mu$s) }
\label{Figure11}
\end{figure}
Fig.~\ref{Figure11} gives the variation of the infidelity with the detuning $\Delta$ for the four-level system.
It is shown that the optimized STIRAP scheme also has periodic fluctuation on the change of the detuning, and the fluctuation is more obvious with the increase of the detuning. This indicates that the larger the detuning $\Delta$ is, the larger the infidelity may be, and there will be some periodic windows of the detuning. In these detuning windows, the infidelity is low, and as the detuning $\Delta$ increases, the low infidelity in these windows will be further reduced. This makes it possible to optimize the STIRAP scheme to achieve adiabatic population transfer in the case of large detuning. The traditional STIRAP scheme has no enough transfer efficiency on the detuning $\Delta$. With the increase of the detuning, the infidelity of the traditional STIRAP increases, which means that the population transfer is suppressed and the influence of nonadiabatic transition is more obvious.

\subsection{F. Population transfer process in the large detuning case}
\begin{figure}[h]
\centering
\includegraphics[width=0.3\textwidth]{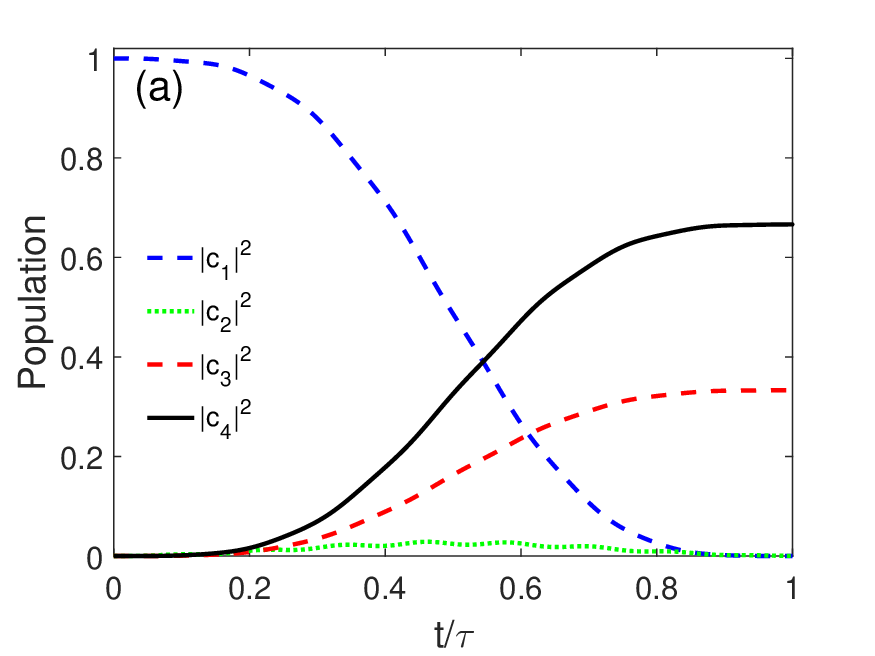}
\includegraphics[width=0.3\textwidth]{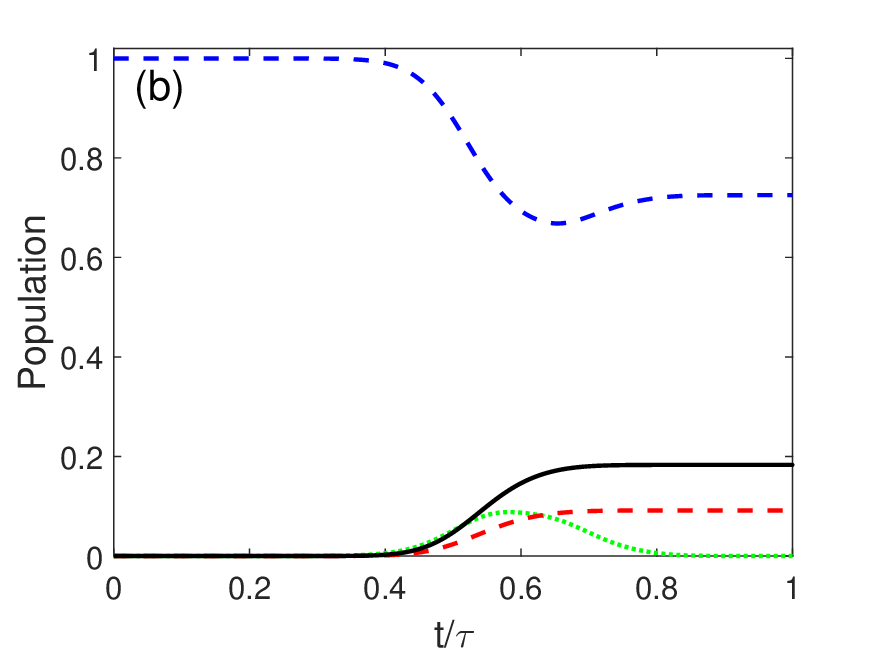}
\caption{Population's evolution of the four-level system for the large detuning ($\Delta>\Omega_0$): (a) The result for the optimal STIRAP scheme; (b) The result for the traditional STIRAP scheme (The blue dashed line, the green dotted line, the red dashed line and the black solid line correspond to the populations in the states $|1\rangle$, $|2\rangle$, $|3\rangle$ and $|4\rangle$, respectively; The pulse operating time $\tau$=7.4$\mu$s, the pulse peak $\Omega_0$=22.13MHz, and the detuning $\Delta$=58.1MHz) }
\label{Figure12}
\end{figure}

From the above discussion, it is not difficult to see that the optimized STIRAP scheme has the ability to achieve the coherent superposition of the final states $\left|3\right\rangle$ and $\left|4\right\rangle$ in the case of short operation time, low pulse peak value, and large detuning $(\Delta >\Omega_0)$, which is another advantage over the traditional STIRAP.
This means that it is possible to prepare quantum superposition states in a specific environment.
For example, the pulse parameter $\chi=\arccos(1/3)$ can be selected for transfer under the conditions of the operation time $\tau=7.4\mu\text{s}$, the pulse peak value $\Omega_0=22.13\text{MHz}$, and the detuning $\Delta = 58.1\text{MHz}$, and the calculating result is shown in Fig.~\ref{Figure12}.

It can be seen that the population transfer in the traditional STIRAP scheme is not complete in this large detuning case.
Most of the population remains in the initial state $\left|1\right\rangle$, while the final states $\left|3\right\rangle$ and $\left|4\right\rangle$ are less populated. In contrast, the optimized STIRAP scheme can achieve complete population transfer even at large detuning. The population in the initial state $\left|1\right\rangle$ becomes very small, and almost all the population is transferred to the states $\left|3\right\rangle$ and $\left|4\right\rangle$, with the final state population satisfying the expected coherent superposition. Therefore, under the conditions where the traditional STIRAP scheme fails to perform well, the optimized STIRAP scheme can still realize efficient adiabatic population transfer.

\subsection{E. Preparation of quantum superposition states in arbitrary proportion}
As mentioned above, only by controlling the optimized pulse parameter $\chi$, the population on the final state can be controlled and the coherent superposition of states in any proportion can be realized.
The following context in this section show the change of the population on the states $\left|3\right\rangle$ and $\left|4\right\rangle$ at the evolution endpoint with respect to the parameter $\chi$.
\begin{figure}[h]
\centering
\includegraphics[width=0.3\textwidth]{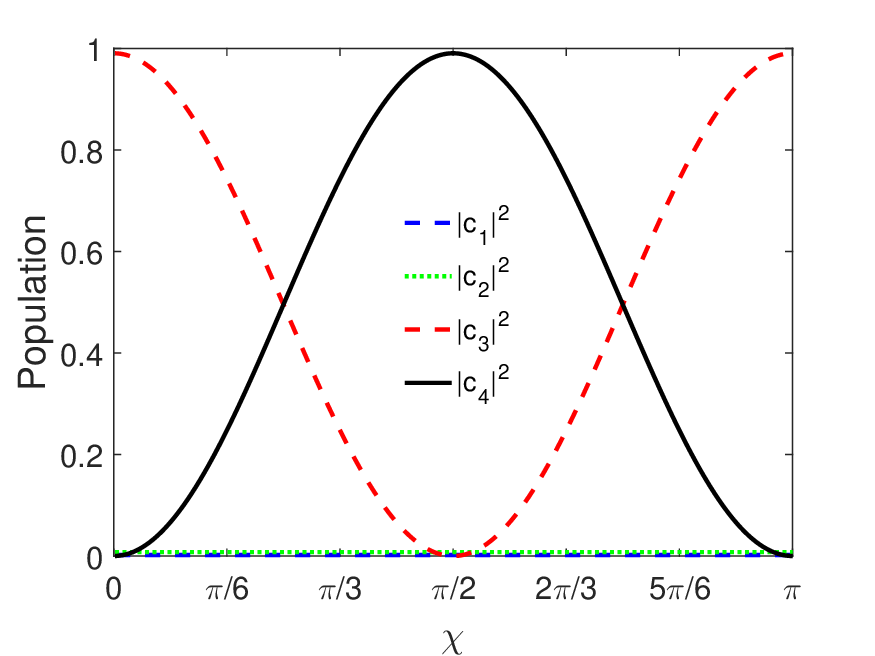}
\caption{Change of the populations for the four-level system with the parameter $\chi$ (The blue dashed line, the green dotted line, the red dashed line and the black solid line correspond to the populations in the states $|1\rangle$, $|2\rangle$, $|3\rangle$ and $|4\rangle$, respectively; The pulse operating time $\tau$=7.4$\mu$s, the pulse peak $\Omega_0$=35MHz, and the detuning $\Delta$=2$\pi$MHz)}
\label{Figure13}
\end{figure}
The variation of the population with $\chi$ for the optimized STIRAP scheme shown in Fig.~\ref{Figure13} brings into correspondence with Eq.~(\ref{Formula13}); that is, the population on the state $\left|3\right\rangle$ is $\cos^2\!\chi$ and the population on the state $\left|4\right\rangle$ is $\sin^2\!\chi$ after the evolution, so that the coherent superposition of the final states can be designed by controlling the pulse parameter $\chi$. Fig.~\ref{Figure14}(a) shows the evolution result of the optimized STIRAP with pulse parameter $\chi = \pi/4$. The particle populations in the states $\left|3\right\rangle$ and $\left|4\right\rangle$ remain equal throughout the evolution, and the $1{:}1$ superposition state is realized. The evolution results of the optimized STIRAP with the pulse parameter $\chi = \arccos(\sqrt{3}/2)$ are given in Fig.~\ref{Figure14}(b), and the $3{:}1$ superposition state is finally realized.
\begin{figure}[h]
\centering
\includegraphics[width=0.3\textwidth]{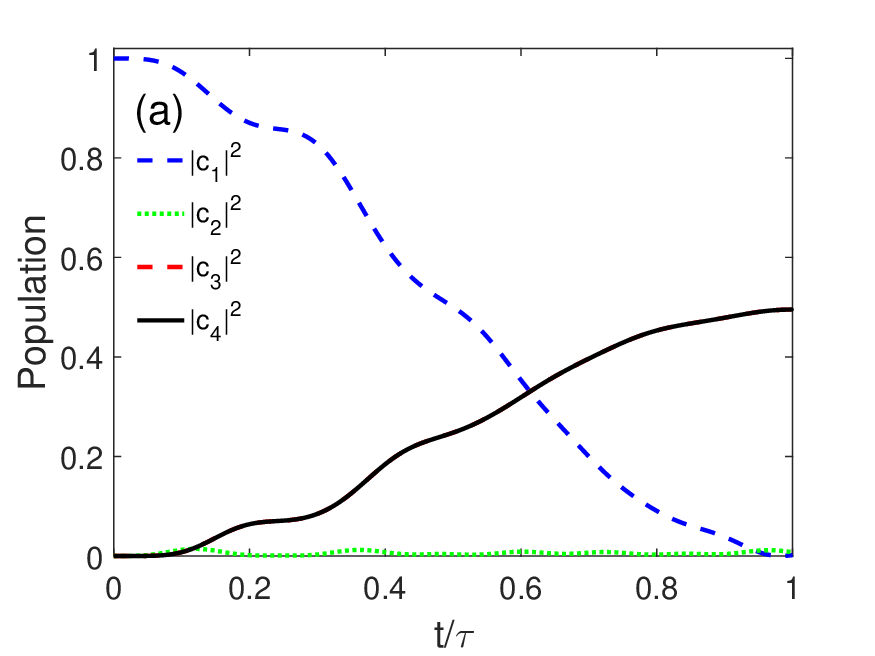}
\includegraphics[width=0.3\textwidth]{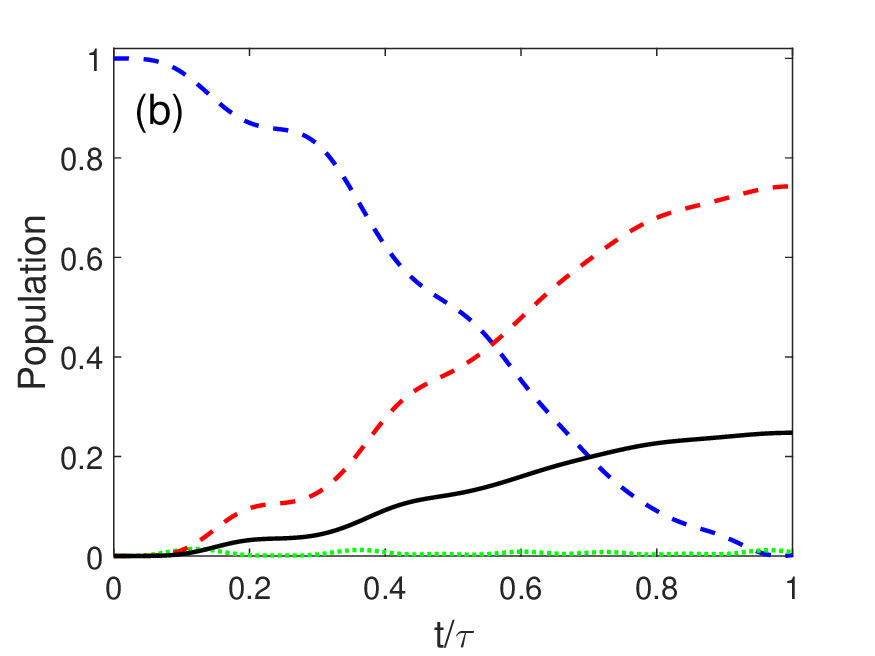}
\caption{Evolution results of the populations: (a) Pulse parameter $\eta$=$\pi/4$; (b) Pulse parameter $\eta$=$\pi/6$ (The blue dashed line, the green dotted line, the red dashed line and the black solid line correspond to the populations in the states $|1\rangle$, $|2\rangle$, $|3\rangle$ and $|4\rangle$, respectively; The pulse operating time $\tau$=7.4$\mu$s, the pulse peak $\Omega_0$=35MHz, and the detuning $\Delta$=2$\pi$MHz) }
\label{Figure14}
\end{figure}

\section*{V. Conclusion }
In this paper, we propose an optimization scheme for the traditional STIRAP based on the dynamical quantum geometric tensor.
The advantages of the optimized STIRAP in adiabatic population transfer are demonstrated by studying a $\Lambda$-type three-level system and a tripod-type four-level system, both with and without single photon detuning. On the one hand, by using optimal control theory and the definition of the quantum geometric tensor, the optimal combination of Rabi pulses is analytically derived such that the total nonadiabatic transition rate of the system remains constant. On the other hand, by comparing the numerical results with those of the Gaussian pulse combination used in the traditional STIRAP, it is found that the optimized STIRAP process exhibits higher transfer efficiency and greater robustness to variations of the pulse duration, pulse amplitude fluctuation, and single photon detuning. The optimized STIRAP process demonstrates the adiabatic resonance as a function of the pulse duration, Rabi pulse amplitude fluctuation, and single-photon detuning. By selecting parameters within the resonant window, high-fidelity population transfer can be achieved with shorter operation time, smaller pulse amplitude, and specific single photon detuning. In particular, for the case of two degenerate dark states in a four-level system, a condition for removing the coupling between degenerate levels is derived in the eigenstate representation. By appropriately choosing the pulse parameters, the optimized STIRAP scheme enables the preparation of quantum superposition states with arbitrary population ratios for the four-level system. In summary, the optimized STIRAP scheme via the dynamical quantum geometric tensor achieves faster and more efficient adiabatic population transfer than the traditional STIRAP scheme. The optimization framework presented in this paper is general and can be extended to study population transfer in other multi-level systems, providing a theoretical foundation for the optimal control of quantum systems.

\end{document}